\documentclass[fleqn,10pt]{wlscirep}
\usepackage[utf8]{inputenc}
\usepackage[T1]{fontenc}
\usepackage{graphicx}
\usepackage{subcaption}
\usepackage{amsmath}
\usepackage{outlines}
\usepackage{listings}

\DeclareMathAlphabet{\mathcal}{OMS}{cmsy}{m}{n}

\usepackage{xcolor}

\title{Calibrating cardiac electrophysiology models using latent Gaussian processes on atrial manifolds}

\author[1,*]{Sam Coveney}
\author[2]{Caroline H Roney}
\author[3]{Cesare Corrado}
\author[4]{Richard D Wilkinson}
\author[5]{Jeremy E Oakley}
\author[3]{Steven A Niederer}
\author[6]{Richard H Clayton} 

\affil[1]{Leeds Institute of Cardiac and Metabolic Medicine, University of Leeds, Leeds LS2 9JT, UK}
\affil[2]{School of Engineering and Materials Science, Queen Mary University of London, London, E1 4NS, UK}
\affil[3]{Division of Imaging Sciences and Biomedical Engineering, King’s College London, London, WC2R 2LS, UK}
\affil[4]{School of Mathematical Sciences, University of Nottingham, Nottingham, NG7 2RD, UK}
\affil[5]{School of Mathematics and Statistics, University of Sheffield, Sheffield, S10 2TN, UK}
\affil[6]{Insigneo Institute for In-Silico Medicine and Department of Computer Science, University of Sheffield, Sheffield, S10 2TN, UK}

\affil[*]{coveney.sam@gmail.com}


\begin{abstract}
Models of electrical excitation and recovery in the heart have become increasingly detailed, but have yet to be used routinely in the clinical setting to guide personalized intervention in patients. One of the main challenges is calibrating models from the limited measurements that can be made in a patient during a standard clinical procedure. 
In this work, we propose a novel framework for the probabilistic calibration of electrophysiology parameters on the left atrium of the heart using local measurements of cardiac excitability. Parameter fields are represented as Gaussian processes on manifolds and are linked to measurements via surrogate functions that map from local parameter values to measurements. The posterior distribution of parameter fields is then obtained.
We show that our method can recover parameter fields used to generate localised synthetic measurements of effective refractory period.
Our methodology is applicable to other measurement types collected with clinical protocols, and more generally for calibration where model parameters vary over a manifold.
\end{abstract}
\begin{document}

\flushbottom
\maketitle

\thispagestyle{empty}

\section*{Introduction}

Mechanical contraction of the heart is initiated and synchronised by a travelling wave of electrical excitation and recovery that arises spontaneously in the natural pacemaker. The heart is made up of four chambers: the ventricles pump blood to the body and lungs, while the atria act as reservoirs and primers for the ventricles. A cardiac arrhythmia is a disturbance of regular heart rhythm resulting in a rapid, slow, or irregular rhythm. Atrial fibrillation (AF) is a common and increasingly prevalent cardiac arrhythmia \cite{staerk_atrial_2017}. AF can be sustained by re-entry, where electrical activation continually propagates into recovering tissue, creating a self-sustaining rotating wave \cite{gray_mechanisms_1995}. Radio-frequency catheter ablation can be used to disrupt re-entrant circuits that act to sustain AF, but is not always effective \cite{Parameswaran2021}.

Two properties of cardiac tissue are important for the development of sustained re-entry, and these properties vary across atrial tissue. Conduction velocity (CV) describes the speed at which an activation wave spreads. The effective refractory period (ERP) is the minimum time interval between two successive stimuli that allows two activation waves to propagate and is related to action potential duration (APD), which is the interval between local activation (depolarization) and recovery (repolarization). Both CV and ERP decrease at shorter pacing intervals, and this dynamic behaviour and its spatial heterogeneity is important for determining the stability of re-entry \cite{qu_cardiac_1999, garfinkel_preventing_2000} as well as the complex paths followed by electrical activation during AF \cite{Loewe2019}. Natural variability in the speed of the excitation wave and the dynamics of excitation and recovery exist both between individuals and within the heart of a single individual \cite{narayan_repolarization_2011, krummen_mechanisms_2012}.
Cardiac tissue exhibits spatial heterogeneity with differences in ion channel conductances, gap junction distributions, and fibrotic remodelling across the heart \cite{cochet_age_2015}. These spatial heterogeneities in structural and functional properties lead to heterogeneity in ERP. The resulting dispersion in repolarisation properties is a mechanism for focal arrhythmia initiation, \cite{bishop_structural_2014} and atrial fibrillation initiation through increasing vulnerability to re-entry \cite{fareh_importance_1998, krummen_mechanisms_2012}.

Electrophysiology (EP) models describe how electrical activation diffuses through cardiac tissue. Local activation and recovery are represented by a set of differential equations describing a reaction-diffusion system that models tissue-scale propagation of activation and cellular activation and recovery \cite{clayton_models_2011, niederer_computational_2019}. Models of cardiac electrical activation have become valuable research tools, but are also beginning to be used in the clinical setting to guide interventions in patients \cite{prakosa_personalized_2018, Boyle2019}. These applications require personalised models of both anatomy and electrophysiology to be constructed. Personalised anatomical models can be assembled from medical images, and statistical shape models enable the assessment of varying shape on electrical behaviour \cite{corrado_quantifying_2020}. Calibration of EP models is difficult because of the limited measurements that can be made routinely in the clinical setting. 

EP model parameters determine model behaviour and for a personalised model should be calibrated to reconstruct the heterogeneity in CV and ERP, as well as their dynamic behaviour, in the heart of a specific patient. Measurements of local activation time (LAT), which measures the time of arrival of the activation wavefront relative to the timing of a pacing stimulus, enable reconstruction of heterogeneous CV for pacing at a fixed rate \cite{roney_technique_2018, puyol_anton_piemap_2021}. 
Calibration to the \emph{dynamics} of activation and recovery is more challenging. Both the quantity and type of data that can be recorded from patients are constrained by the clinical procedure, so it is difficult to determine spatial heterogeneity of repolarisation. An S1S2 pacing protocol can be used to measure restitution curves. The heart is paced for several beats at an initial pacing cycle length S1, followed by a stimulus with a shorter length S1. This protocol is repeated for different values of the S2 interval, and the shortest S2 that can elicit an activation indicates an upper bound for ERP at the stimulus site.
While models can be calibrated to reconstruct CV(S2) restitution and ERP from LAT measurements with an S1S2 protocol \cite{corrado_personalized_2017, corrado_work_2018}, recent work raises doubts over whether model parameters can be identified uniquely from these types of measurement alone \cite{coveney_bayesian_2021}. Biophysically detailed models of electrical activation have large numbers of parameters and many of these may be unidentifiable from restitution curve data \cite{whittaker_calibration_2020}. There is a need for robust approaches that can interrogate cardiac tissue properties more thoroughly while at the same time minimising additional interventions.

In this paper, we present a novel method for probabilistic calibration of an electrophysiology simulator from spatially sparse measurements using a probabilistic model of electrophysiology parameters on a manifold representing the left atrium of the heart. We focus on estimating parameter fields that reconstruct heterogeneity in ERP. We chose to use a phenomenological EP model that captures the main features of cardiac activation and recovery.
We determine two types of ERP measurements for calibrating EP parameters that determine excitability. EP parameters are modelled as latent Gaussian processes (GPs) on a manifold, and linked to observations via surrogate functions and a likelihood function designed for ERP measurements. We use Markov Chain Monte Carlo (MCMC) to obtain the posterior distribution of EP parameter fields across the atrium. We validate our method quantitatively by generating ground truths and calibrating to sparse data. The principles behind our method generalise to other measurement types, such as CV and APD restitution data, making our approach a step forward in the creation of digital twins capable of reproducing the complex dynamics of electrophysiology.

\section*{Results}

\subsection*{Workflow}

The computational model, or `simulator', that we seek to calibrate is composed of (i) a finite element mesh representing an atrial manifold $\mathbf{x} \in \Omega$; 
(ii) an electrophysiology model, that maps EP parameters $\theta_l(\mathbf{x}), l=1,2,\ldots$ defined on the computational mesh to observable quantities; and (iii) a numerical solver for running EP simulations.
Details on obtaining and processing a mesh for suitability in electrophysiology simulations are given in Methods, including details on the example mesh used here. The EP model is the modified Mitchell-Schaeffer (mMS) model \cite{mitchell_two-current_2003, corrado_two-variable_2016}, the parameters of which are effectively time-constants representing different phases of the action potential. We parameterize the mMS model with the following 5 parameters: $CV_{max}(\mathbf{x}), \tau_{in}(\mathbf{x}), \tau_{out}(\mathbf{x}), \tau_{open}(\mathbf{x}), APD_{max}(\mathbf{x})$. See Methods for details on the simulation model, parameterisation, and allowable parameter ranges, and details on the numerical implementation. We use the software \emph{openCARP} \cite{plank_opencarp_2021} to solve the mono-domain model for our simulations.

\begin{figure}
    \centering
    \includegraphics{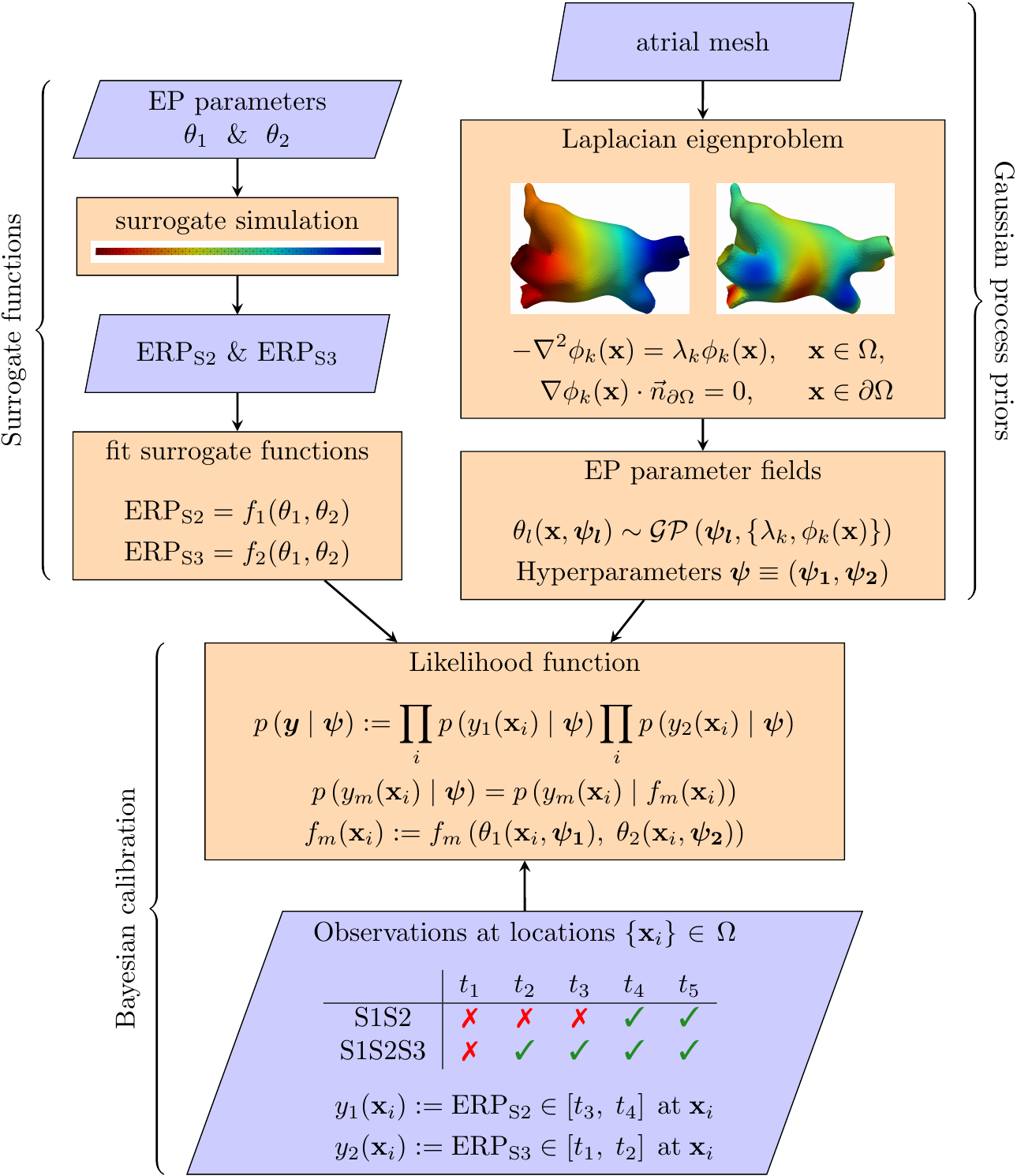}
    \caption{Workflow for using ERP observations to calibrate model parameters. A simplified `surrogate simulation' consisting of a strip of tissue paced from one end is used to determine ERP values for a design of parameters. Surrogate functions are fit to these data, allowing rapid prediction of ERP from model parameters. Electrophysiology parameter fields are modelled as Gaussian processes on the atrial manifold, using a reduced-rank formulation relying on eigenfunctions and eigenvalues of the Laplace-Beltrami operator on the atrial mesh. ERP measurements using an S1S2 (or S1S2S3) protocol measure whether successful activation occurs for S2 (or S3) times $t_1$ < $t_2$ < $t_3 \ldots$ etc.
    Using a likelihood function designed for ERP measurements and the surrogate functions, the likelihood can be evaluated given hyperparameters for the GP models. This allows for probabilistic calibration by obtaining the posterior distribution of hyperparameters using MCMC.}
    \label{fig:workflow}
\end{figure}

The main task in this work is to calibrate the simulator by inferring parameter fields $\theta_l(\mathbf{x})$ from ERP measurements. Figure \ref{fig:workflow} represents our modeling workflow, which we summarize here. Our code is available in a Zenodo repository \cite{sam_coveney_2022_7081857}.

\textbf{Surrogate functions:} The simulator can be used to map parameter fields to ERP fields $\text{ERP}(\mathbf{x}) = \text{sim}(\theta_l(\mathbf{x}), l=1,2,\ldots)$. Given ERP observations at multiple locations on the atrial mesh, as well as an appropriate likelihood model for these observations, the simulator could be used in a MCMC setting in order to calibrate the parameter fields by obtaining samples from the posterior distribution for the EP fields. However, this is an extremely inefficient approach, since ERP depends only on local (rather than remote) tissue properties. We utilize a surrogate function (also called an "emulator") solution in which we learn the mapping from parameters to ERP. This surrogate function allows us to predict ERP at location $\mathbf{x}$ as $\text{ERP}(\mathbf{x}) = f(\theta_l(\mathbf{x}), l=1,2,\ldots)$, bypassing the need to run the simulator directly for inference.

\textbf{Gaussian process priors:} The mesh has $\approx 10^5$ vertices for which parameters need to be defined, but ERP measurements are restricted to a subset of these vertices, with number of observations on the order of $10^0$--$10^1$. The electrophysiology parameter fields must be assumed to have low-rank structure, induced by spatial correlation, in order to make inferences about EP parameter values at locations other than ERP observation locations. This is achieved here by modeling the EP parameters using latent Gaussian process (GP) priors $\theta_l(\mathbf{x}) \sim \mathcal{GP}$. We use Gaussian Process Manifold Interpolation (GPMI), a method  we proposed for defining Gaussian process  (GP) distributions on manifolds \cite{coveney_gaussian_2020}. The approach  uses solutions $\left\{\lambda_k, \phi_k(\mathbf{x})\right\}$ of the Laplacian (Laplace-Beltrami) eigenproblem on the mesh \cite{solin_hilbert_2020}.

\textbf{Bayesian calibration:} We perform probabilistic calibration with MCMC to obtain the posterior distribution of latent variables in the GPs. We utilize a likelihood function that we developed specifically for ERP measurements, which accounts for how an S1S2 pacing protocol to determine ERP effectively measures the S2 interval in which ERP lies, rather than measuring ERP directly.

\subsection*{Sensitivity analysis and surrogate functions}

\begin{figure}
\centering
\begin{subfigure}{1.0\textwidth}
    \centering
    \includegraphics[width=0.7\textwidth]{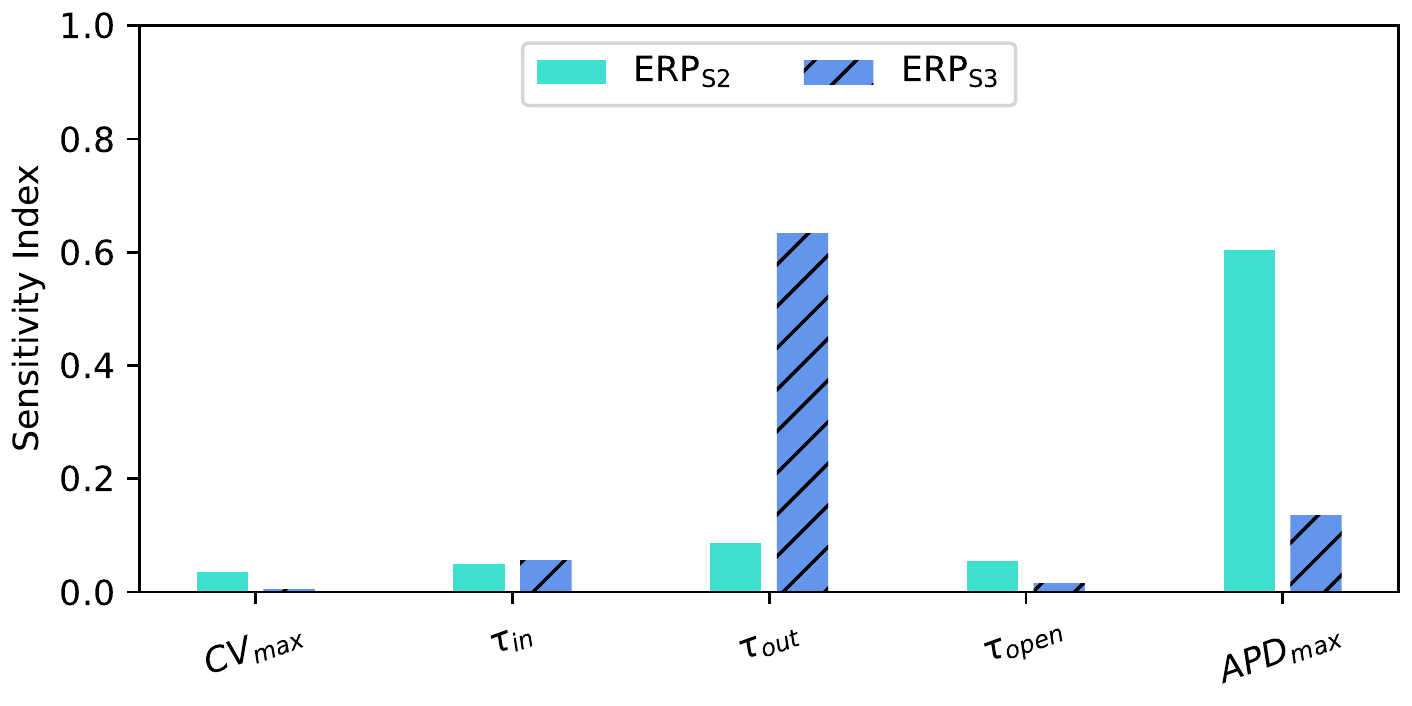}
    \caption{Sensitivity analysis of $\text{ERP}_\text{S2}$ and $\text{ERP}_\text{S3}$ for mMS parameters.} \label{fig:sensitivity}
\end{subfigure}
\begin{subfigure}{1.0\textwidth}
    \centering
    \includegraphics[width=0.8\textwidth]{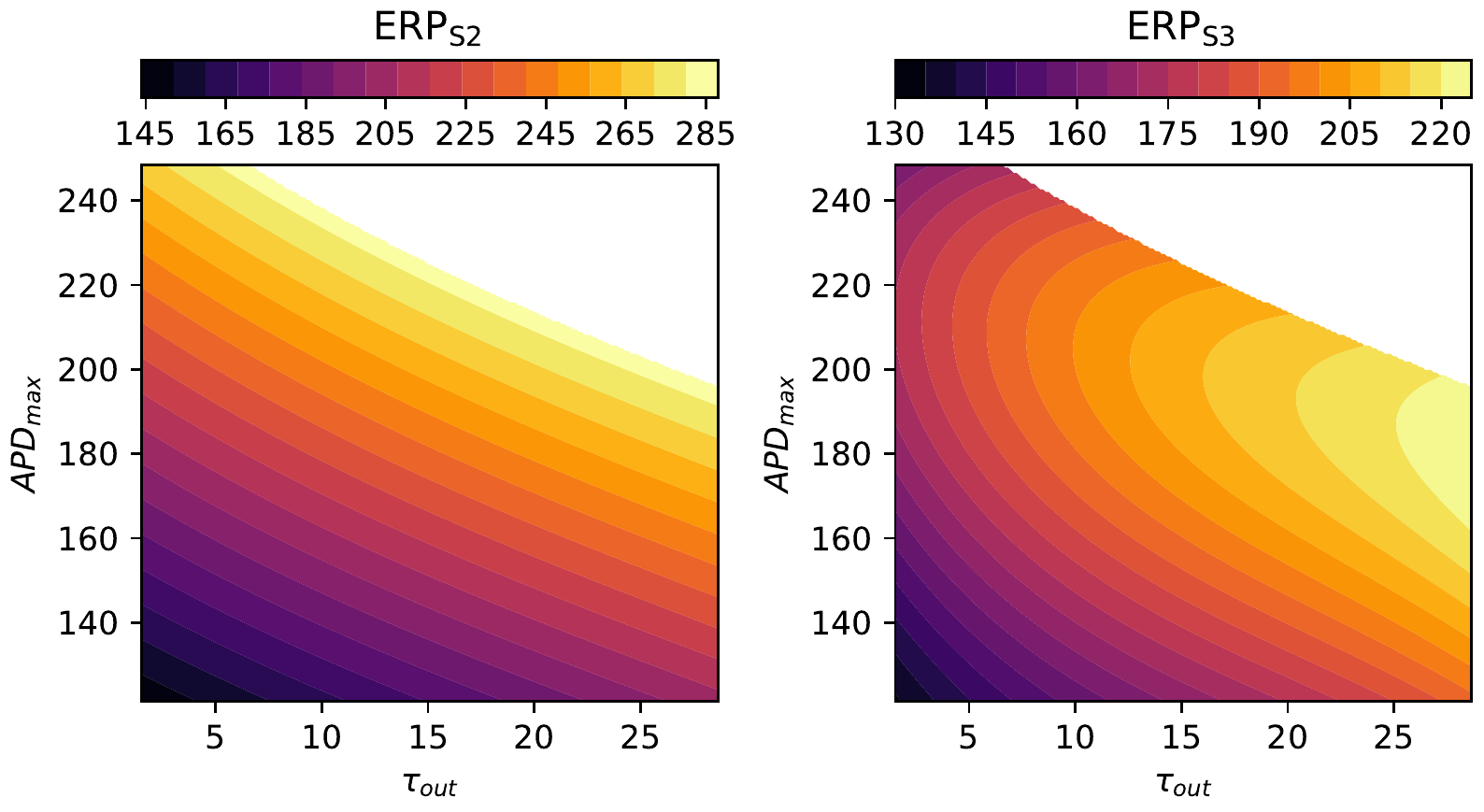}
    \caption{Surrogate functions of $\text{ERP}_\text{S2}$ and $\text{ERP}_\text{S3}$ as a function of $\tau_{out}$ and $APD_{max}$.}  \label{fig:surrogates}
\end{subfigure}
    \caption{\textbf{(a)} Sensitivity analysis of ERP for the mMS electrophysiology model for all parameters. $\tau_{out}$ and $APD_{max}$ have the largest sensitivity indices for both types of ERP. \textbf{(b)} Surrogate functions for predicting ERP from $\tau_{out}$ and $APD_{max}$. A discontinuity in $\text{ERP}_\text{S3}$ occurs at $\text{ERP}_\text{S2} \approx 285$~ms, so this region is shown in white.}
\end{figure}

Figure \ref{fig:sensitivity} shows sensitivity indices for two types of ERP: an ERP measurement for S1S2 with S1 600~ms, denoted here as $\text{ERP}_\text{S2}$ and another type of ERP measurement for S1S2S3 pacing for S1 600~ms S2 300~ms, denoted here as $\text{ERP}_\text{S3}$. The S1S2S3 protocol, consisting of N S1 beats, 1 S2 beat, and 1 S3 beat, is introduced in this paper. We have determined that these measurements can be used to calibrate EP parameters sufficiently to reproduce not only these ERP measurements, but also the time required for the action potential to reach various levels of repolarization recovery (e.g. $\text{APD}_{20}$ and $\text{APD}_{90}$, the time required for 20\% and 90\% recovery).  It is a key finding that the S1S2S3 protocol can be used (alongside the standard S1S2 protocol) to disentangle the contributions of separate parts of the action potential to the value of ERP, without needing to measure the action potential directly.

The sensitivity indices in Figure \ref{fig:sensitivity} show that these ERP measurements are mainly determined by $\tau_{out}$ and $APD_{max}$, which approximately correspond to the duration of the repolarization and plateau phases of the action potential respectively. Calibration of other parameters, which determine some aspects of the shape of restitution curves but which do not strongly impact ERP, require both CV and APD restitution curve data from an S1S2 protocol \cite{coveney_bayesian_2021}. For this reason, we have determined to use ERP to calibrate $\theta_1 \equiv \tau_{out}$ and $\theta_2 \equiv APD_{max}$. Figure \ref{fig:surrogates} shows contour plots of the surrogate functions for ERP. A discontinuity occurs in the $\text{ERP}_\text{S3}$ surface for parameters combinations resulting in $\text{ERP}_\text{S2} > 285$~ms, so data for $\text{ERP}_\text{S2} > 280$~ms were discarded before fitting this function. Note that the majority of clinical $\text{ERP}_\text{S2}$ measurements fall in the range $170-270$~ms, so even $280$~ms could be considered as an upper limit \cite{bode_repolarization-excitability_2001}.

\subsection*{Synthetic experiments} %

To test our methodology, we ran synthetic experiments as detailed in Methods. We used a left atrial mesh generated from a scan of an individual performed at St Thomas’ Hospital (see methods for details). We created ground truth parameter fields for $\tau_{out}$ and $APD_{max}$ in order to verify our calibration approach. We used 10 measurement locations, placed at random using a maximin design that excluded sites close to the mesh boundaries. The resolution of the S1S2 and S1S2S3 protocol was set to 10~ms. We used 24 eigenfunctions for representing each of the two parameter fields in Eq. \eqref{eq:GP}, which we found to be sufficient to capture spatial variation while allowing good posterior sampling. For MCMC we used 5000 iterations, using 8 chains, discarding the first 50\% of the samples as `burn-in', and randomly thinning the remaining samples by a factor of 100 to give 200 posterior samples.

\begin{figure}
\centering
\begin{subfigure}{0.49\textwidth}
    \centering
    \includegraphics[width=\textwidth]{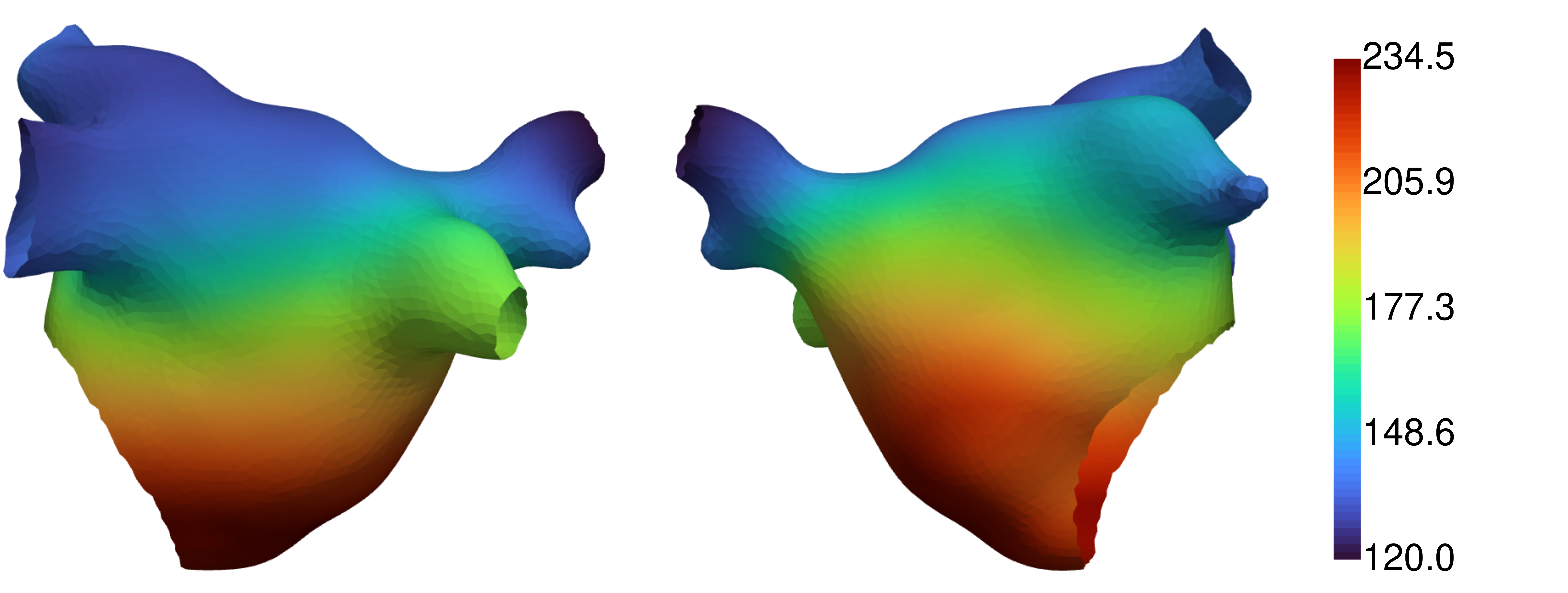}
    \caption{$APD_{max}$ true values.} \label{fig:APDmax_true}
\end{subfigure}%
\hspace{1.0pt}
\begin{subfigure}{0.49\textwidth}
    \centering
    \includegraphics[width=\textwidth]{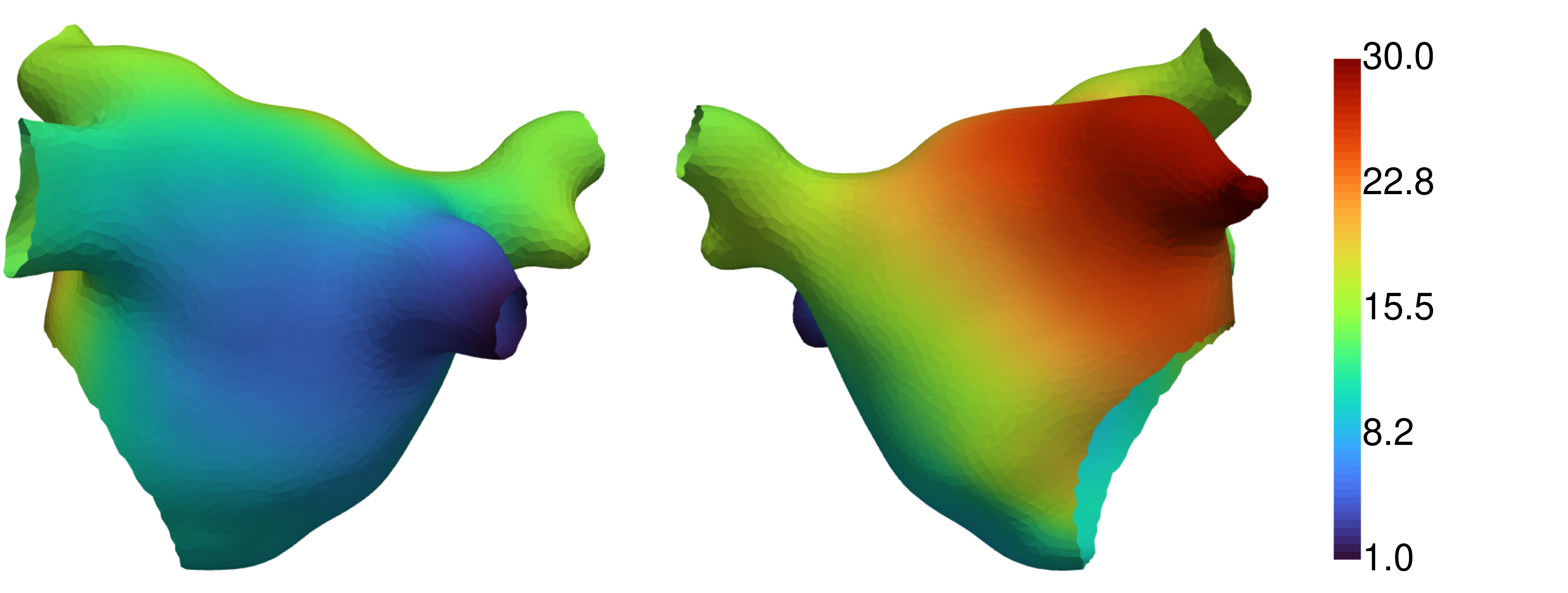}
    \caption{$\tau_{out}$ true values.} \label{fig:tauout_true}
\end{subfigure}%
\\
\begin{subfigure}{.49\textwidth}
    \centering
    \includegraphics[width=\textwidth]{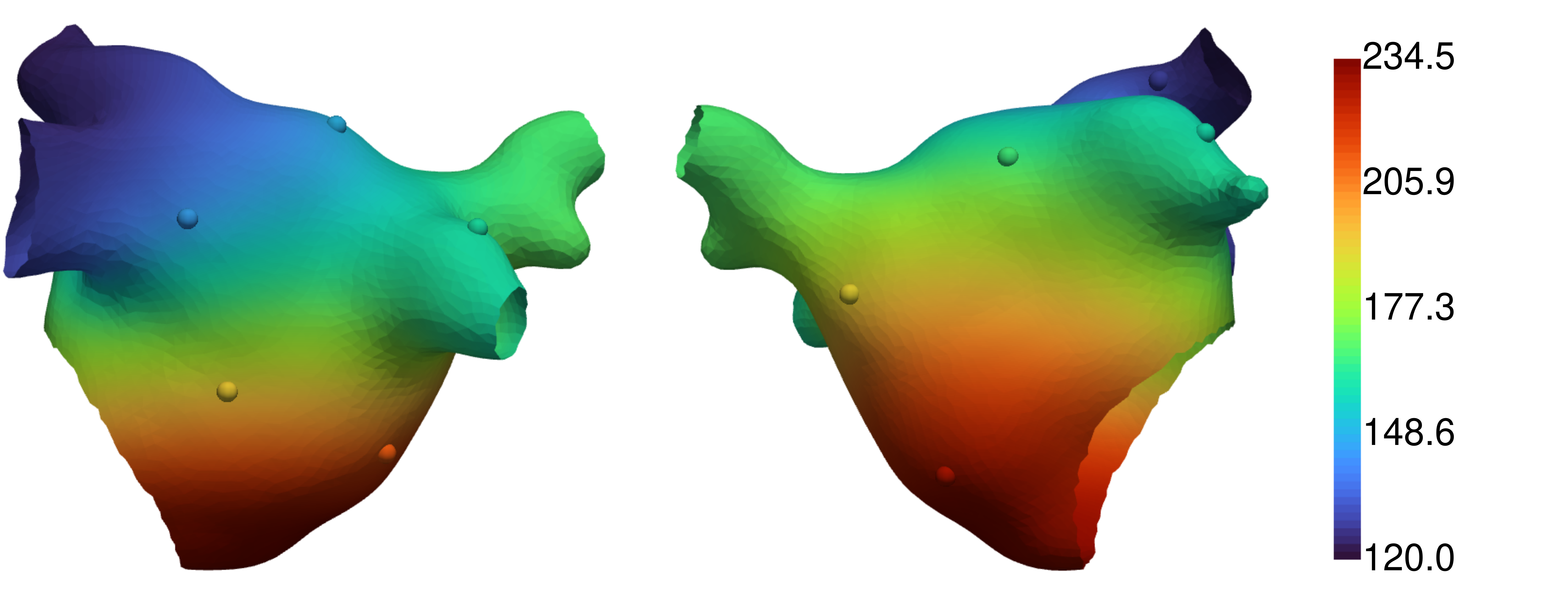}
    \caption{$APD_{max}$ posterior mean.} \label{fig:APDmax_mean}
\end{subfigure}%
\hspace{1.0pt}
\begin{subfigure}{.49\textwidth}
    \centering
    \includegraphics[width=\textwidth]{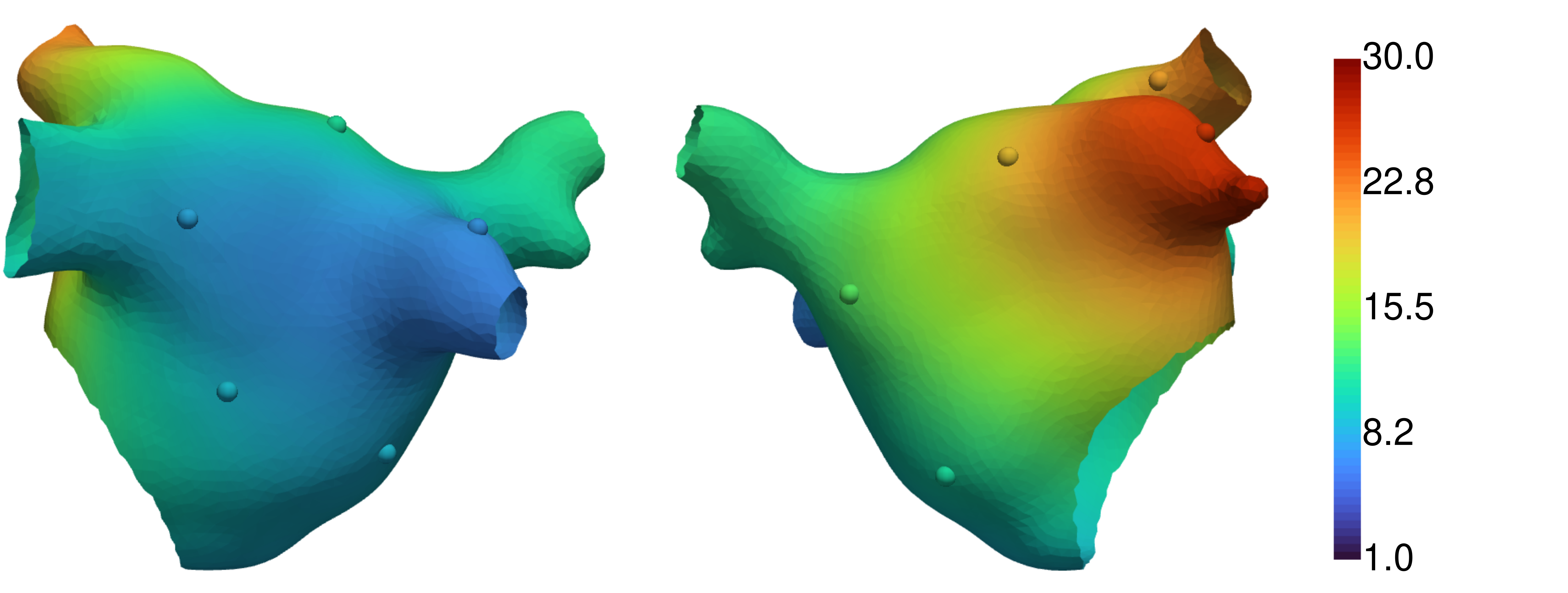}
    \caption{$\tau_{out}$ posterior mean.} \label{fig:tauout_mean}
\end{subfigure}%
\\
\begin{subfigure}{.49\textwidth}
    \centering
    \includegraphics[width=\textwidth]{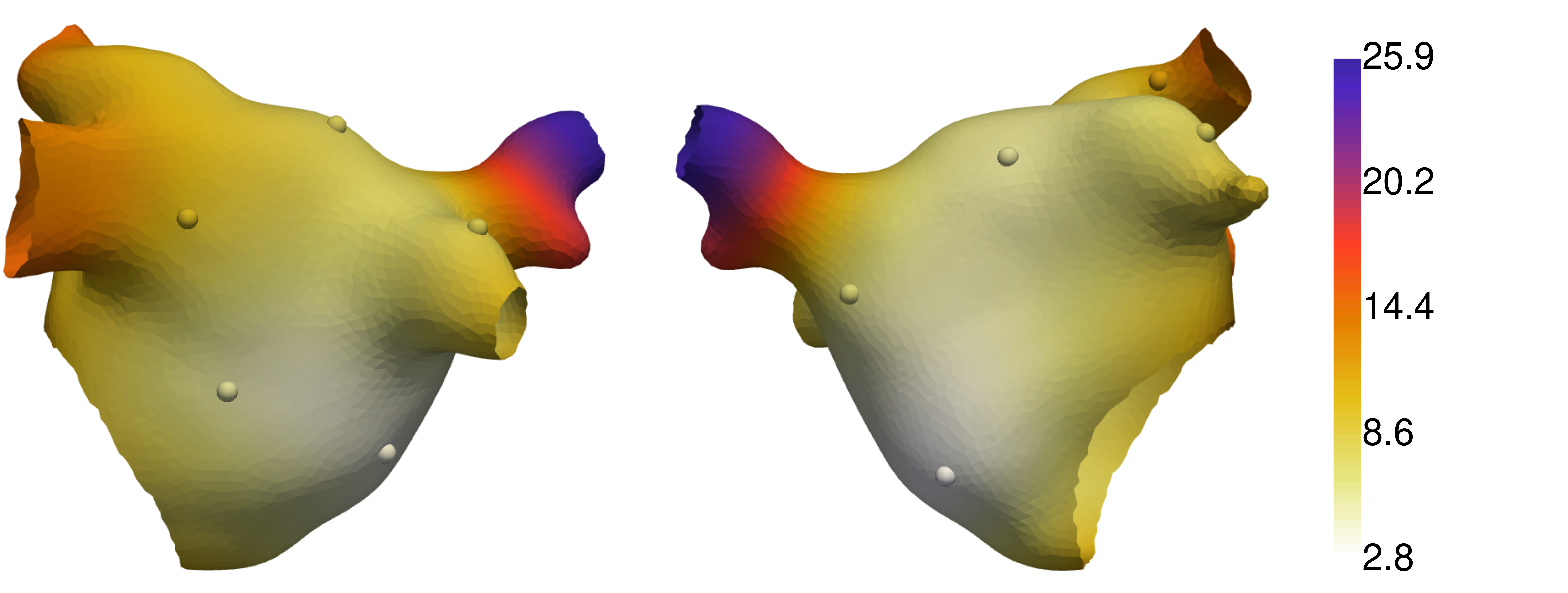}
    \caption{$APD_{max}$ posterior standard deviation.}\label{fig:APDmax_stdev}
\end{subfigure}%
\hspace{1.0pt}
\begin{subfigure}{.49\textwidth}
    \centering
    \includegraphics[width=\textwidth]{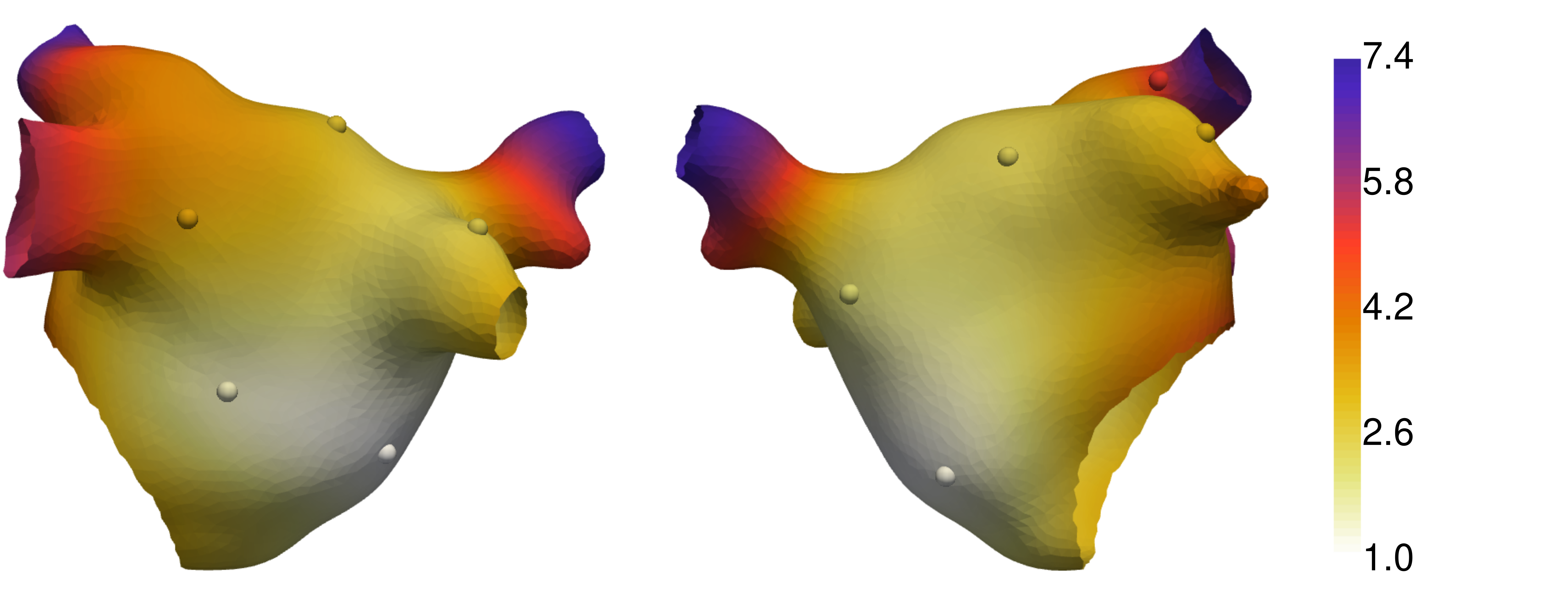}
    \caption{$\tau_{out}$ posterior standard deviation.}\label{fig:tauout_stdev}
\end{subfigure}%
\caption[short]{Ground truth electrophysiology parameter fields and corresponding ERP values and APD values. All units are milliseconds.} \label{fig:parameters}
\end{figure}

\begin{figure}
\centering
\begin{subfigure}{.49\textwidth}
    \centering
    \includegraphics[width=\textwidth]{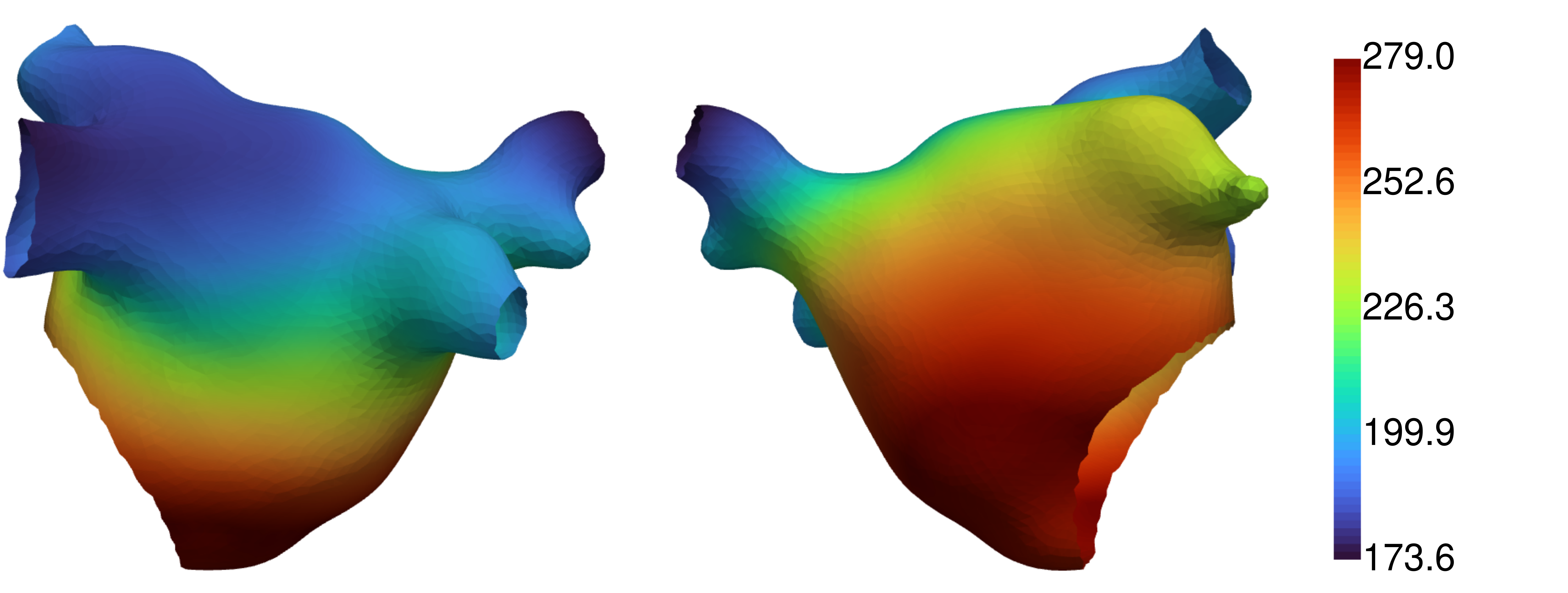}
    \caption{$\text{ERP}_\text{S2}$ true values.}\label{fig:ERPS1_true}
\end{subfigure}
\hspace{1.0pt}
\begin{subfigure}{.49\textwidth}
    \centering
    \includegraphics[width=\textwidth]{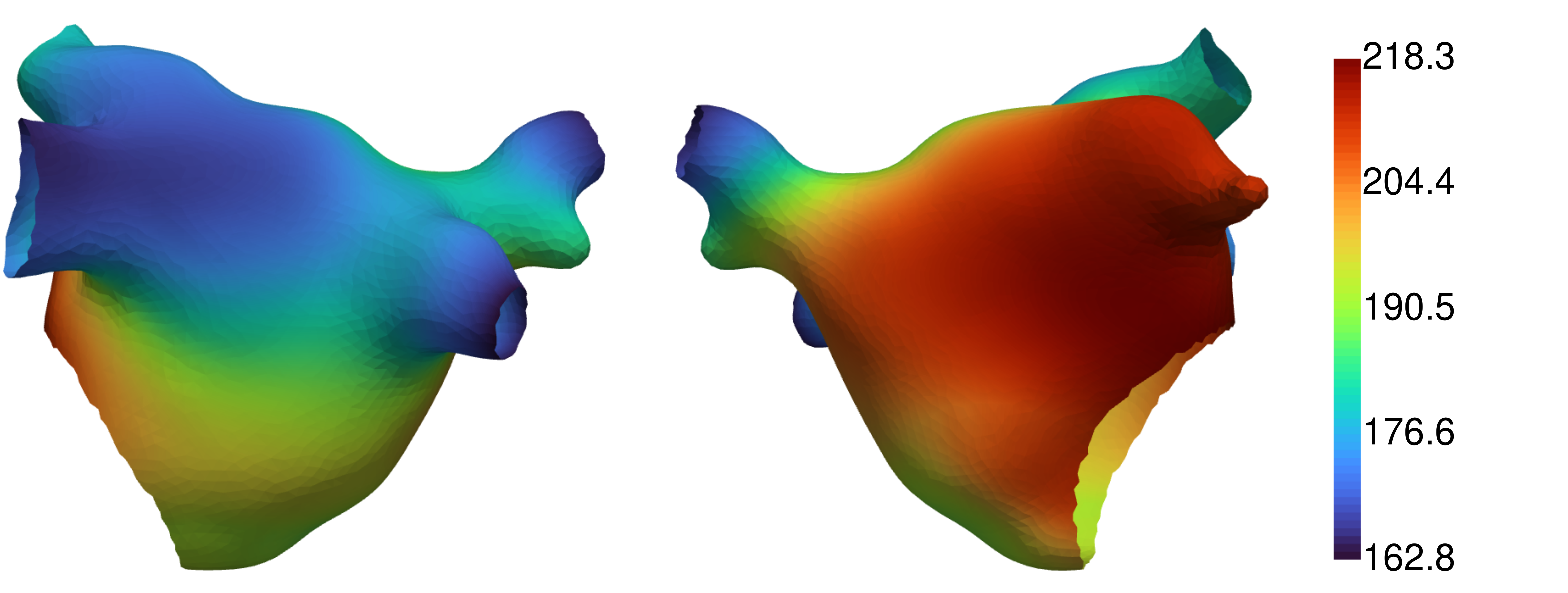}
    \caption{$\text{ERP}_\text{S3}$ true values.}\label{fig:ERPS2_true}
\end{subfigure}%
\\
\begin{subfigure}{.49\textwidth}
    \centering
    \includegraphics[width=1.0\textwidth]{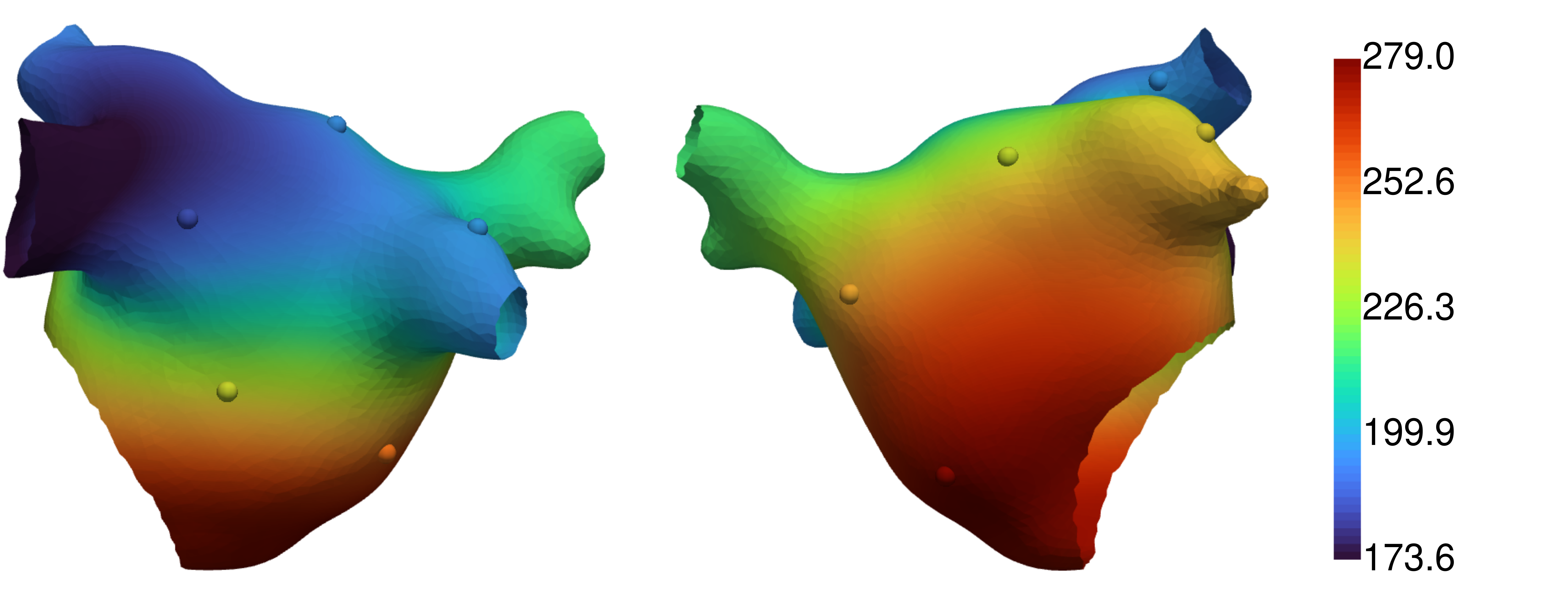}
    \caption{Mean of $\text{ERP}_\text{S2}$ samples.}
    \label{fig:ERP_S1_samples_mean}
\end{subfigure}%
\hspace{1.0pt}
\begin{subfigure}{.49\textwidth}
    \centering
    \includegraphics[width=1.0\textwidth]{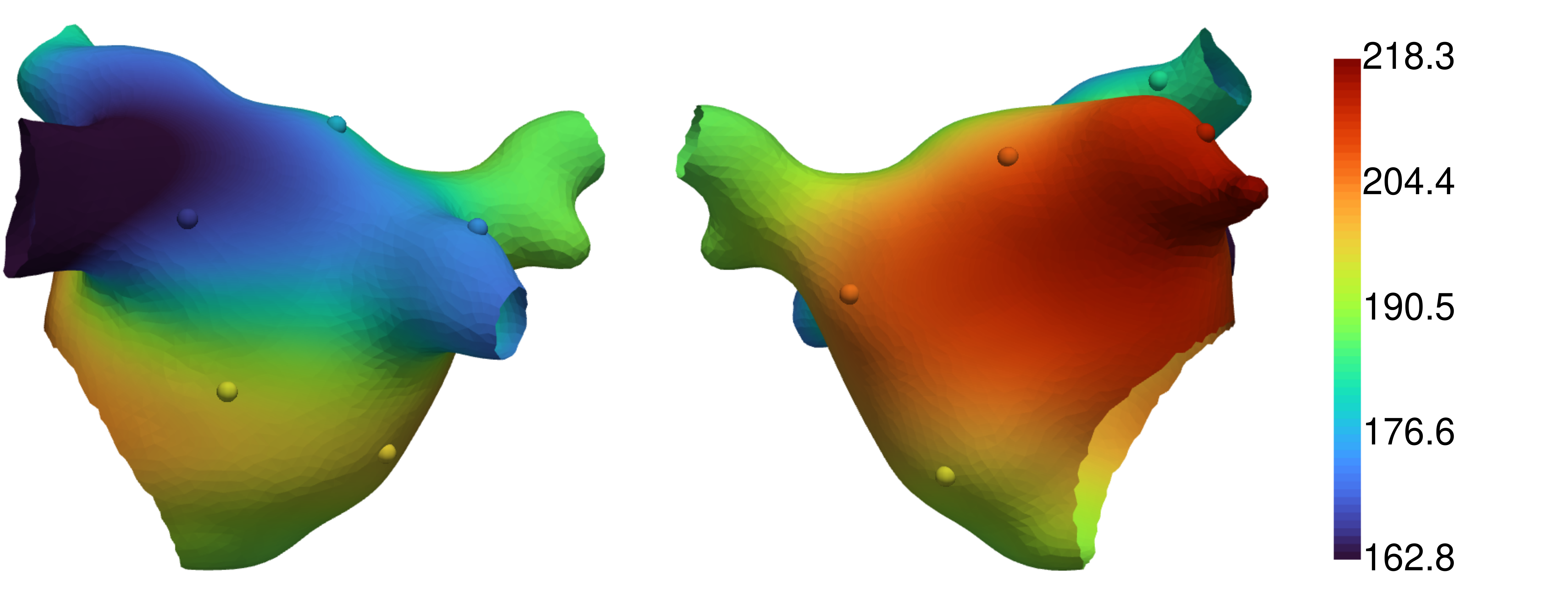}
    \caption{Mean of $\text{ERP}_\text{S3}$ samples.}
    \label{fig:ERP_S1S2_samples_mean}
\end{subfigure}%
\\
\begin{subfigure}{.49\textwidth}
    \centering
    \includegraphics[width=1.0\textwidth]{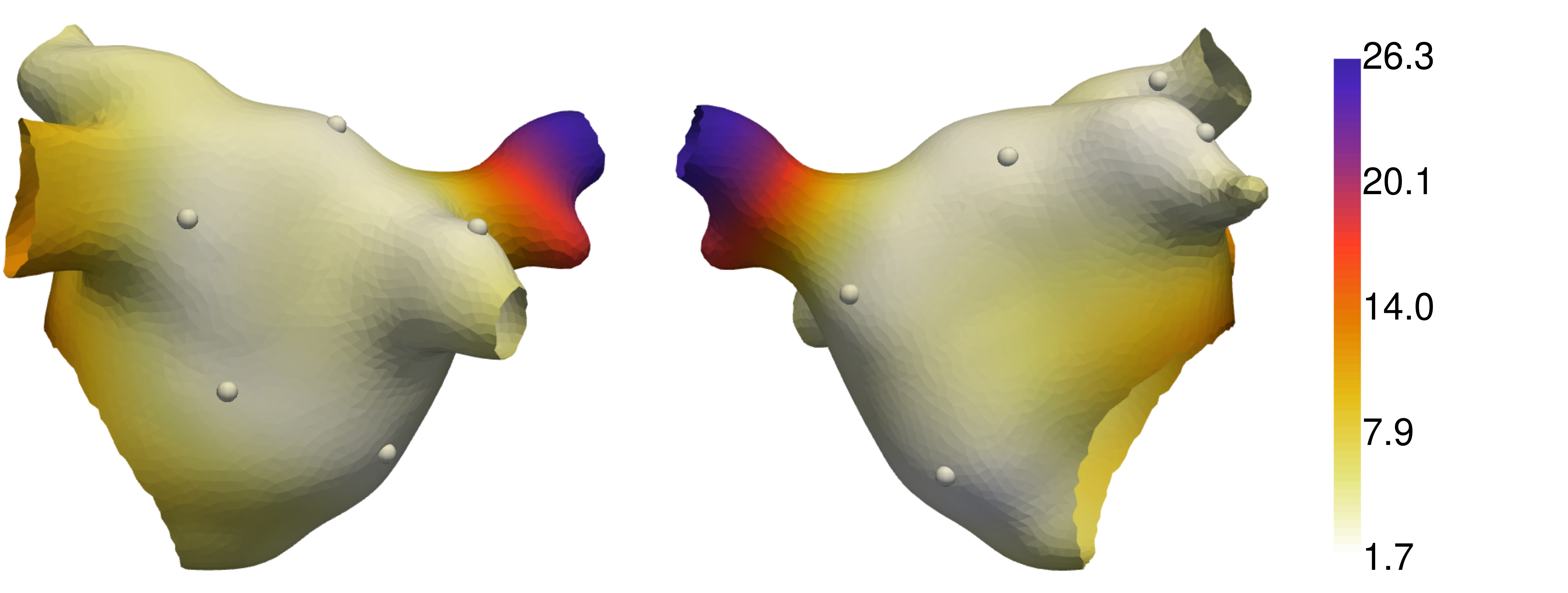}
    \caption{Standard deviation of $\text{ERP}_\text{S2}$ samples.} \label{fig:ERP_S1_samples_std}
\end{subfigure}%
\hspace{1.0pt}
\begin{subfigure}{.49\textwidth}
    \centering
    \includegraphics[width=1.0\textwidth]{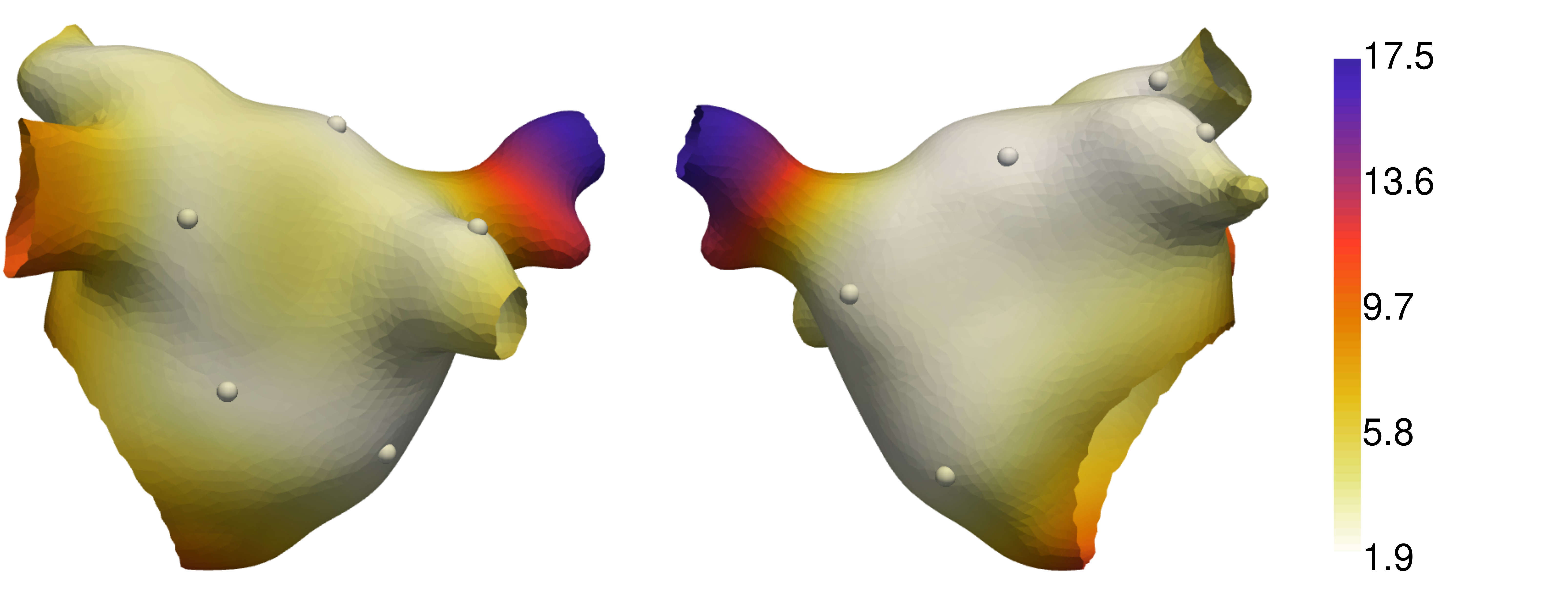}
    \caption{Standard deviation of $\text{ERP}_\text{S3}$ samples.} \label{fig:ERP_S1S2_samples_std}
\end{subfigure}
\begin{subfigure}{.49\textwidth}
    \centering
    \includegraphics[width=1.0\textwidth]{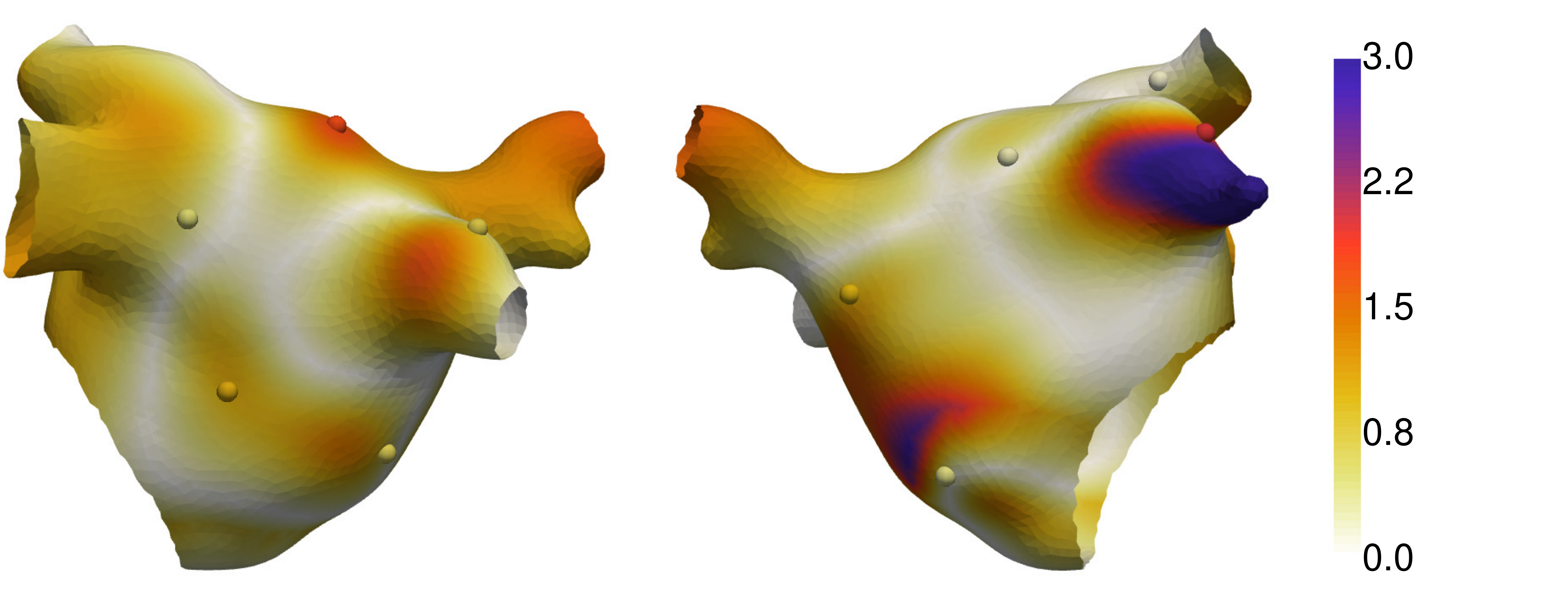}
    \caption{ISE of $\text{ERP}_\text{S2}$ samples.} \label{fig:ERP_S1_samples_ise}
\end{subfigure}%
\hspace{1.0pt}
\begin{subfigure}{.49\textwidth}
    \centering
    \includegraphics[width=1.0\textwidth]{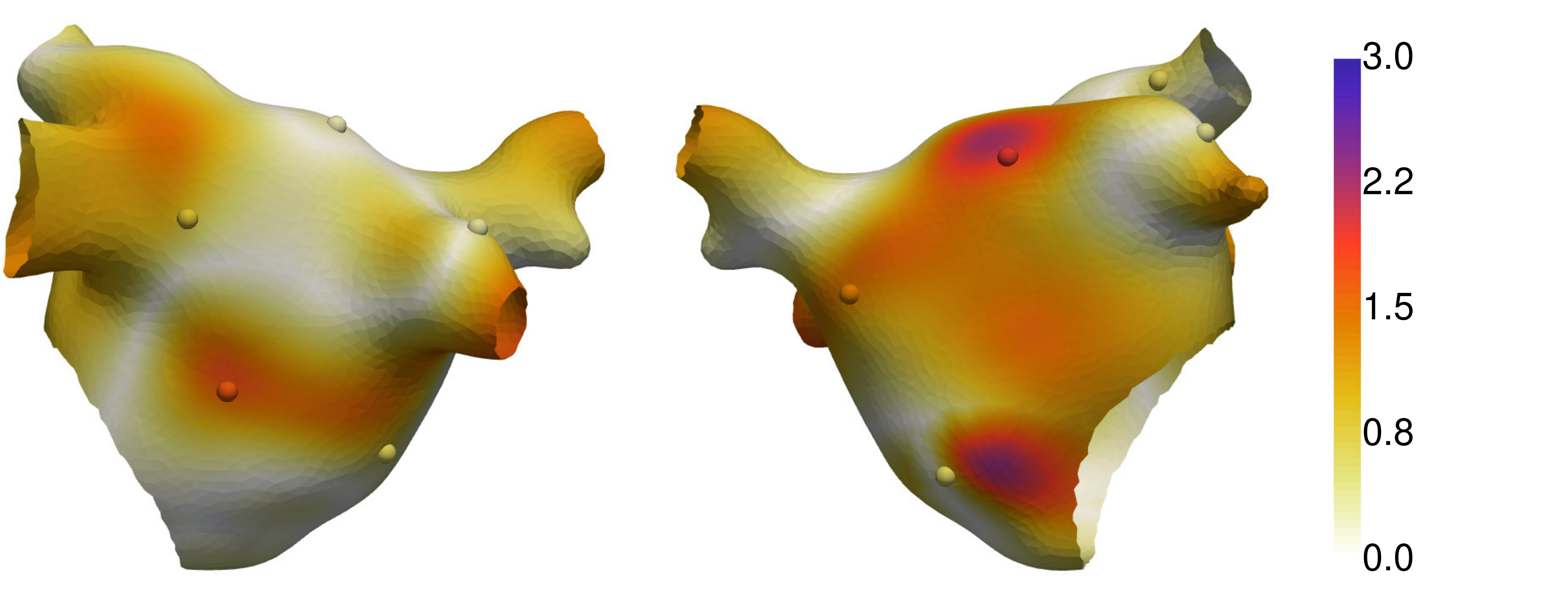}
    \caption{ISE of $\text{ERP}_\text{S3}$ samples.}
    \label{fig:ERP_S1S2_samples_ise}
\end{subfigure}
\caption[short]{Predicted electrophyiology parameter fields (posterior mean) and corresponding ERP values (mean of ERP field samples) and APD values (from posterior mean of parameter fields). The spheres show the location of ERP measurements. All units are milliseconds.} \label{fig:ERP}
\end{figure}

\begin{figure}
\centering
\begin{subfigure}{.49\textwidth}
    \centering
    \includegraphics[width=\textwidth]{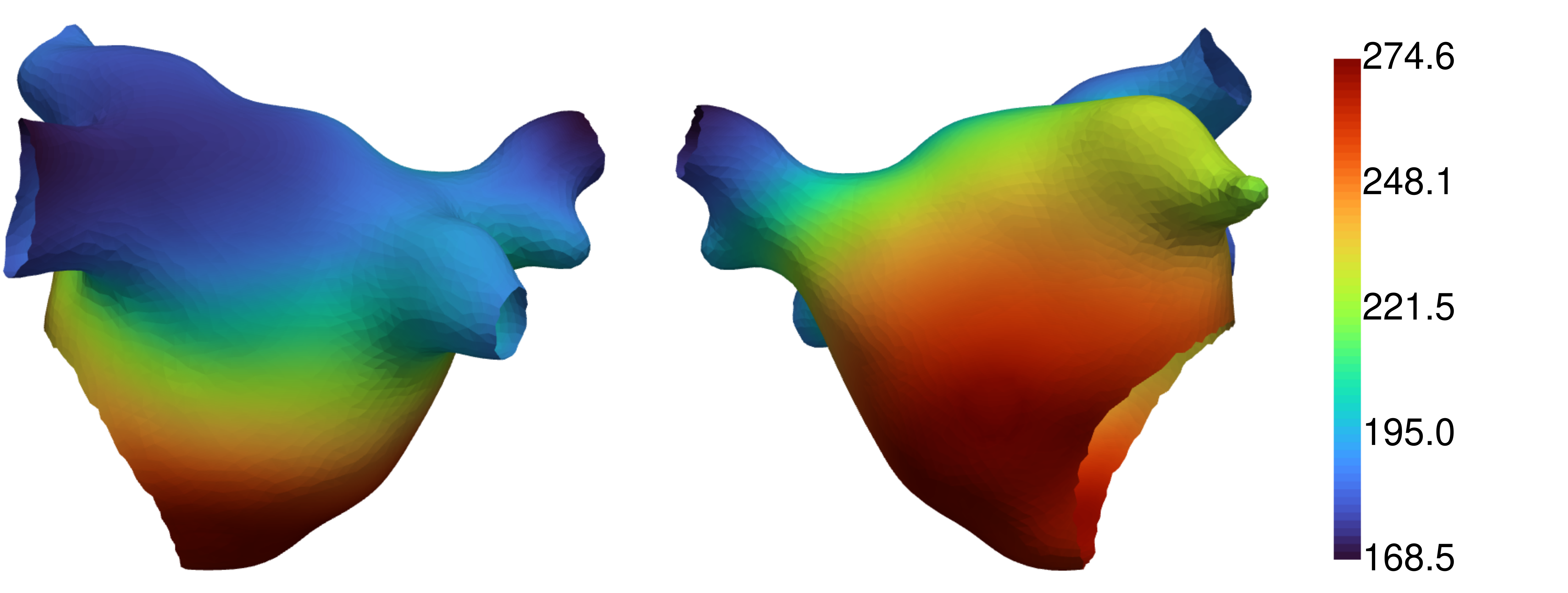}
    \caption{$\text{APD}_{90}$ simulation with true parameters.} \label{fig:APD90_true}
\end{subfigure}%
\hspace{1.0pt}
\begin{subfigure}{.49\textwidth}
    \centering
    \includegraphics[width=\textwidth]{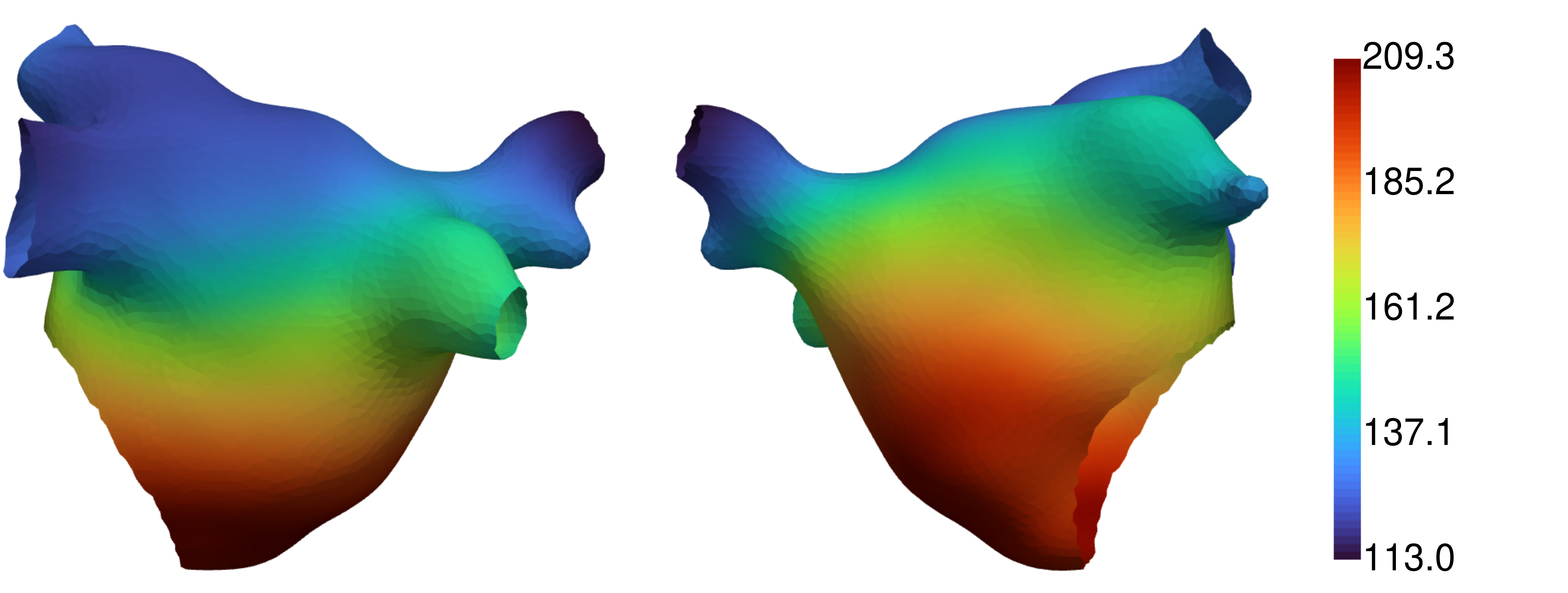}
    \caption{$\text{APD}_{20}$ simulation with true parameters.} \label{fig:APD20_true}
\end{subfigure}%
\\
\begin{subfigure}{.49\textwidth}
    \centering
    \includegraphics[width=\textwidth]{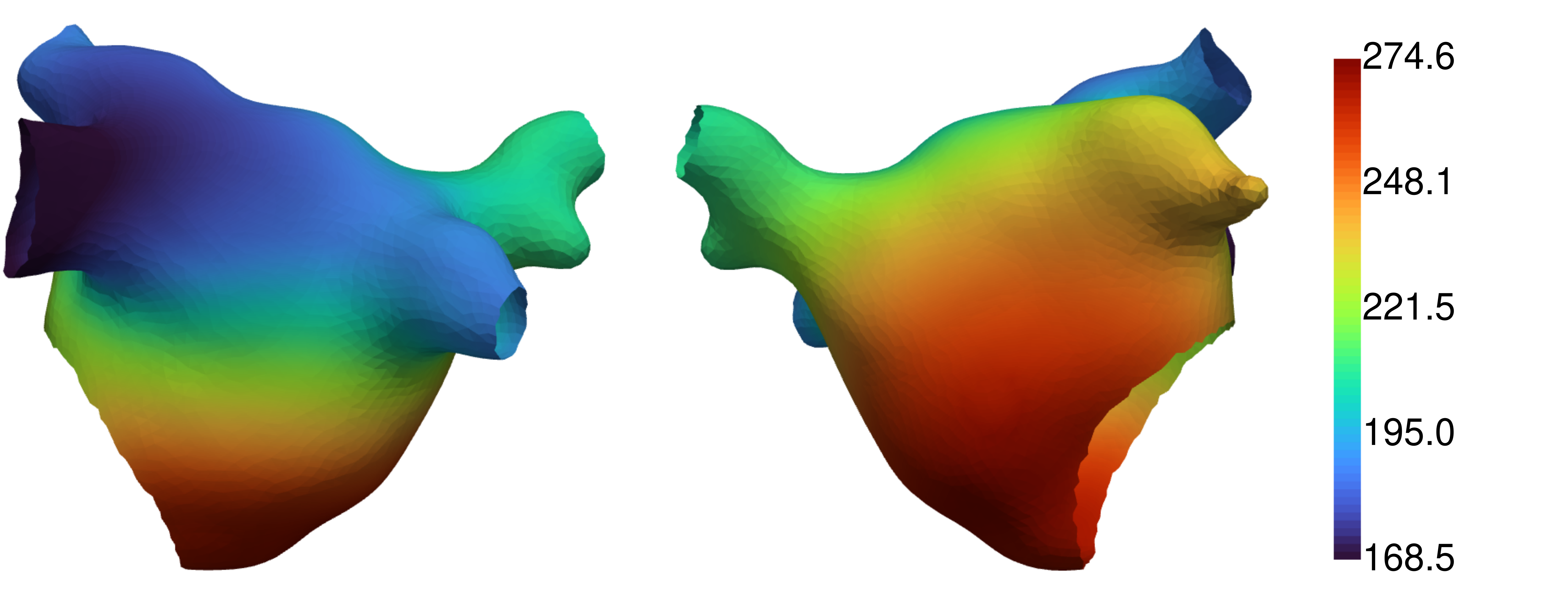}
    \caption{$\text{APD}_{90}$ simulation with predicted parameters.} \label{fig:APD90_pred}
\end{subfigure}%
\hspace{1.0pt}
\begin{subfigure}{.49\textwidth}
    \centering
    \includegraphics[width=\textwidth]{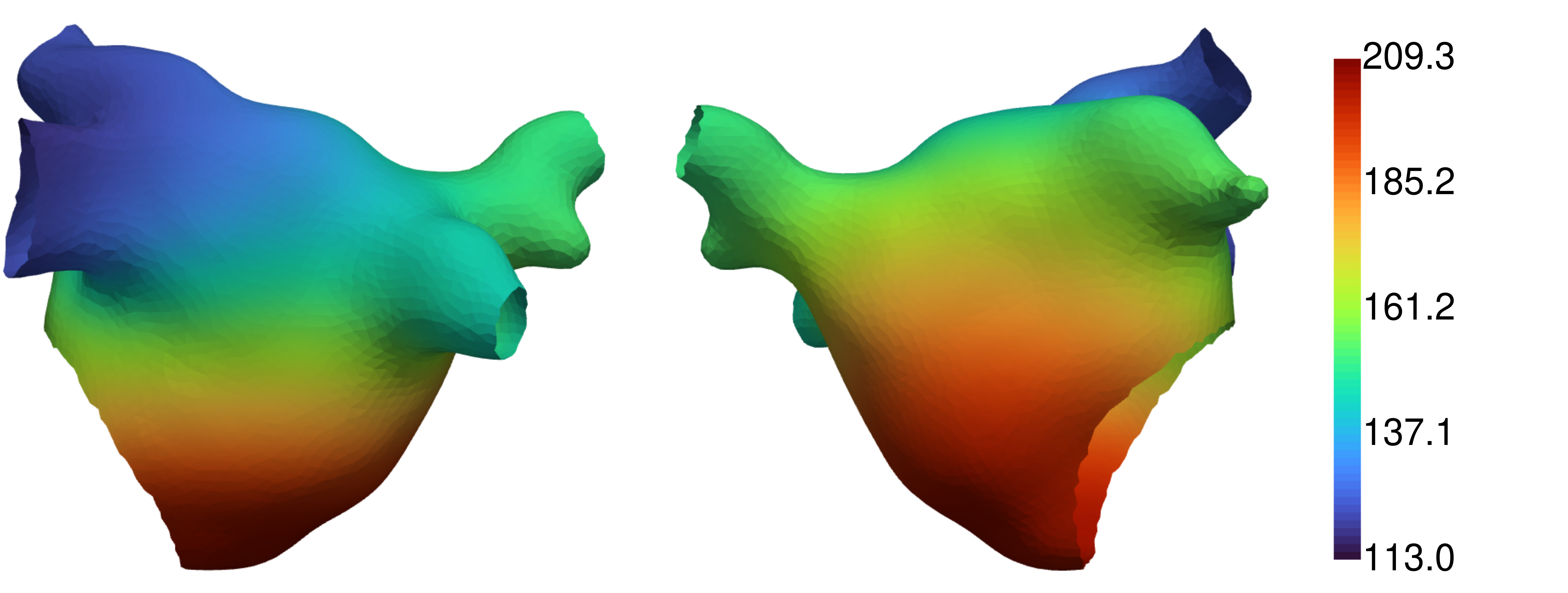}
    \caption{$\text{APD}_{20}$ simulation with predicted parameters.} \label{fig:APD20_pred}
\end{subfigure}%
\\
\caption[short]{Uncertainty of predicted electrophyiology parameter fields (posterior standard deviation) and corresponding ERP values (standard deviation of ERP field samples), and independent standard error (ISE) plots comparing predicted and ground-truth ERP. The spheres show the location of ERP measurements. All units are milliseconds.} \label{fig:APD}
\end{figure}

Figure \ref{fig:parameters} shows the true parameter fields, and the posterior mean and standard deviation of the calibrated parameter fields. Figure \ref{fig:ERP} shows the true ERP fields, the posterior mean and standard deviation of the ERP fields (calculated from ERP samples, which are calculated from the parameter field posterior samples), and the Independent Standard Errors (ISE) of ERP (the absolute difference between true and posterior mean, divided by the posterior standard deviation). Measurement locations are shown as spheres in Figures \ref{fig:parameters} and \ref{fig:ERP}, colored by the corresponding values at each location. Figure \ref{fig:APD} shows the APD simulation results from the atrial simulator using the ground truth parameter fields and the posterior mean of the calibrated parameter fields.

The prediction of EP parameter fields $\tau_{out}$ and $APD_{max}$ and $\text{ERP}_\text{S2}$ and $\text{ERP}_\text{S3}$ captures the ground truth extremely well. Predictions on the pulmonary veins, which are effectively regions of extrapolation, deviate from the ground truth more than other regions on the main body of the atrium. These deviations are on the order of of the S2 and S3 resolution, and the posterior variance is higher in these regions. Uncertainty increases with distance from the measurement locations. The ISE scores show that the \emph{distribution} of ERP predicted by the model covers the ground truth well as nearly all values are less than 3. The ISE for $\text{ERP}_\text{S2}$ on the left atrial appendage is above 3, which may be caused by a combination of high ground-truth values for $\tau_{out}$ (which are not effectively probed by the measurements) and insufficient basis functions to capture high spatial variation in this region of the mesh. APD from the full atrial simulator using the posterior mean of the parameters (the maximum \emph{a posteriori} estimate could have been used instead) matches the ground truth values very closely, demonstrating that the action potential has been calibrated well using only ERP measurements.

\begin{figure}
    \centering
    \includegraphics[width=0.85\textwidth]{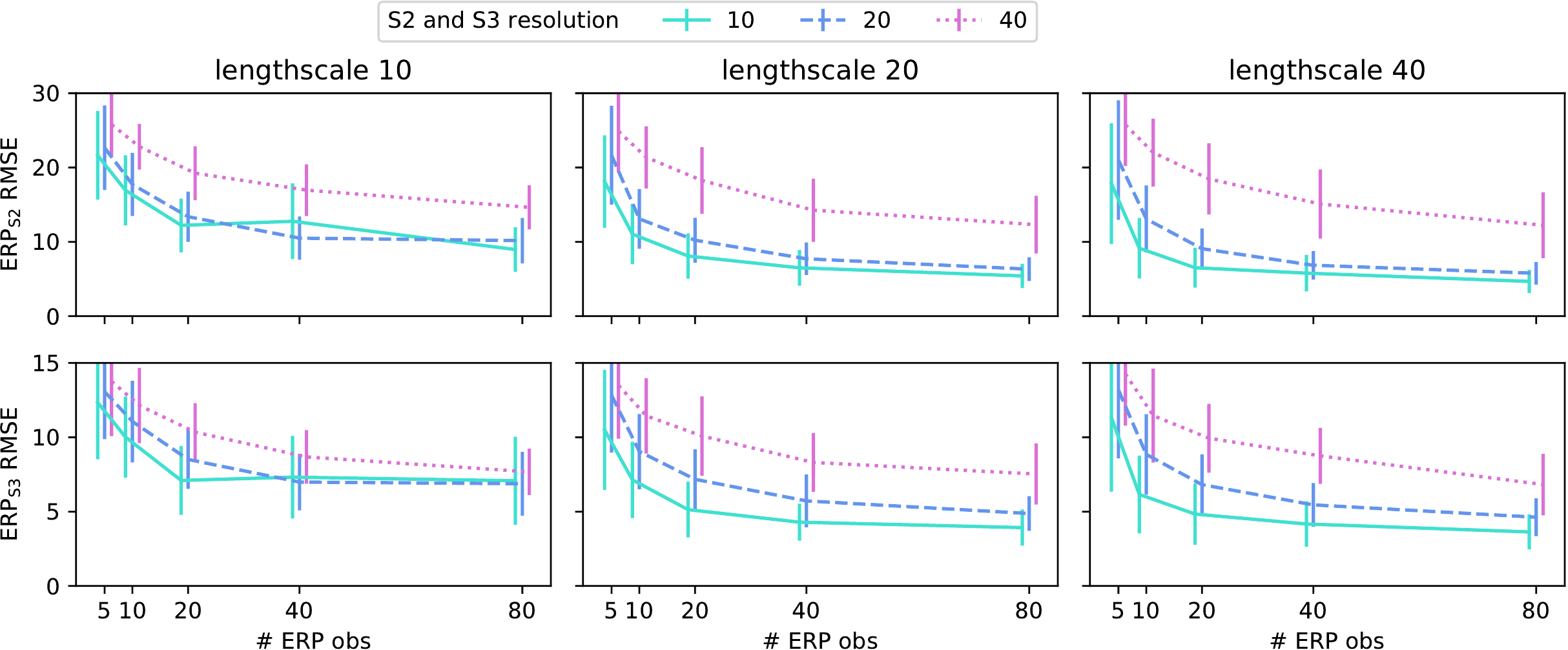}
    \caption{Quantitative validation for ERP prediction, for different lengthscales (1 unit kernel lengthscale $\approx 3.2$~mm), number of observations, and S2 and S3 resolution. The model prediction is from the maximum likelihood estimate. 32 basis functions are used for both EP parameter fields. For each combination of lengthscale, observations, and S2 and S3 resolution, 45 different samples of EP parameter fields are generated, and 5 different observation designs are used for each sample. The RMSE scores of ERP for each validation run is calculated over all mesh vertices. The error bars are 1 standard deviation.} \label{fig:ERP_val}
\end{figure}

We also performed quantitative validation across a broad range of designs. Figure \ref{fig:ERP_val} shows these validation results, for different configurations of the S1S2(S3) pacing protocol (number of ERP observations, resolution of S2 and S3 intervals) and different heterogeneity for ERP, controlled by different correlation lengthscales for generated $APD_{max}$ and $\tau_{out}$ ground-truth fields. A unit of kernel lengthscale is approximately 3.2~mm for this mesh; see Methods for details. Prediction values of ERP are based on the maximum \emph{a posteriori} estimate of the parameters, and here we use 32 eigenfunctions per EP parameter field in order to better model fields with more rapid spatial variation. Root Mean Squared Error (RMSE) is reduced with increasing lengthscale (less ERP heterogeneity), decreasing S2 and S3 resolution (more precise measurements), and increasing number of observations. We note that our likelihood function introduces a small amount of bias, discussed below, which for S2 and S3 resolution 10ms causes RMSE to increase slightly from 20 to 40 observations. Overall, the quantitative validation suggests that little is gained above 20 observation locations.

\section*{Discussion} %

In this paper, we have developed a workflow for calibrating an electrophysiology simulator from sparse measurements of excitability. This was done by representing the spatially varying parameter fields as Gaussian processes on a manifold, and linking these parameters to excitability observations through non-linear surrogate functions (emulators). Using a likelihood function for ERP observations, we performed probabilistic calibration to obtain the posterior distribution of the EP parameter fields. Both visual and quantitative comparison demonstrates that this workflow can successfully calibrate a simulator to ERP to a high level of accuracy. 

The nature of ERP observations, in which only the interval containing ERP is observed (and the possible brackets around this interval are fixed by the S1S2(S3) protocol), is that the ability to learn more by adding observations is strongly limited above a certain point. Figure \ref{fig:ERP_val} demonstrates that this limit is reached faster for smaller S2 and S3 resolution. Our likelihood function does introduce a very small amount of bias, since the true likelihood should be constant in the pacing interval, but our approximation decreases on approaching the interval edges. A simple solution would be to pad the ERP observation brackets, which would remove the bias but reduce the precision. Without the assumption that measurements at locations give information about quantities at nearby locations, i.e. spatial correlation, inference about tissue properties beyond measurement sites would not be possible and atrial tissue would need to be sampled everywhere. Such regularization might make it difficult to capture discontinuous changes in tissue properties, although it would be difficult to measure such abrupt changes in tissue behaviour using sparse measurements. It may be possible to utilize other personal data (e.g. scans) or prior information (e.g. a database of clinical measurements) to assist with calibration.

The latent Gaussian process model serves two purposes. Firstly, a run of the electrophysiology model requires specification of parameters at all points on the mesh, and the Gaussian process enables this specification via interpolation between measurement locations. Secondly, we assume that parameter values at neighbouring locations on the mesh are likely to be similar, which means that we need to do joint inference for the parameters at the measurement locations, rather than inferring parameters at each measurement location independently. In developing our method, we first attempted such an independent inference approach, in which parameters are calibrated at each measurement location independently and then interpolated over the manifold using GPMI, but we were not able to obtain satisfactory results. Our current workflow easily allows more complex spatial modeling using multiple latent GPs per EP parameter field, each with independent covariance kernels and hyperparameters that can be freely given suitable priors. It also provides the benefit of being able to constrain the posterior distribution by directly manipulating the posterior samples based on \emph{a priori} knowledge, such that parameter values (or the tissue properties depending on these parameters) should fall within a certain physiological range.

Our proposed workflow for calibration is suitable for other types of data. We have previously shown that Gaussian processes can be used as surrogate functions for CV, APD, and ERP restitution curves \cite{coveney_bayesian_2021}. Observations from these restitution curves at different locations over the atrium could be included in calibration simply by including additional contributions to the likelihood function and using "Restitution Curve Emulators" to map from EP parameters to the corresponding restitution curves. Our approach here solves the problem of representing the EP parameter fields on a manifold so as to make probabilistic calibration to sparse measurements into a tractable problem. This allows for propagating uncertainty from measurements through to an ensemble of calibrated models.

\section*{Methods} %

\subsection*{Electrophysiology model}

The modified Mitchell-Schaeffer (mMS) cell model \cite{mitchell_two-current_2003, corrado_two-variable_2016} for mono-domain tissue simulations with isotropic diffusion is expressed in the following equations:

\begin{gather}
    \frac{\partial V_m}{\partial t} = D \nabla^2 V_m + 
    h \frac{V_m(V_m - V_{gate})(1 - V_m)}{\tau_{in}} -
    (1 - h) \frac{V_m}{\tau_{out}} + J_{stim} \label{eq:mms_V} \\
    \frac{\partial h}{\partial t} = 
    \begin{cases}
        (1-h) / \tau_{open} & \text {if $V_m \leq V_{gate}$} \\
         -h / \tau_{close} & \text {otherwise}
    \end{cases} \label{eq:mms_h}
\end{gather}
\noindent
where $V_m$ is a normalised membrane voltage, $h$ is a gating parameter that controls recovery, and $J_{stim}$ is an externally applied stimulus. The 4 cell model parameters $\boldsymbol{\tau} = (\tau_{in}, \tau_{close}, \tau_{out}, \tau_{open})$ are time-constants that approximately characterize stages of the action potential sequence, and $D$ is conductivity. We fixed the excitation threshold $V_{gate}$ to 0.1. As in \cite{coveney_bayesian_2021}, we reparameterized the model as follows:

\begin{gather}
    CV_{max} = 0.5(1 - 2 V_{gate}) \sqrt{2D/\tau_{in}}\\
    APD_{max} = \tau_{close} \log\left(1 + \tau_{out} (1 - V_{gate})^2 / (4 \tau_{in})\right)
\end{gather} \label{eq:transformed}

In this new parameter space, weighted combinations of valid parameters are also valid parameters, which means that spatial interpolation of valid parameters will produce valid parameters. We refer to these \emph{transformed} parameters simply as `parameters'. The valid ranges of these parameters are set as $CV_{max}$ 0.1--1.5~m/s, $\tau_{in}$ 0.01--0.30~ms, $\tau_{out}$ 1--30~ms, $\tau_{open}$ 65--215~ms, $APD_{max}$ 120--270~ms.

\subsection*{Atrial Mesh}

To generate the mesh for the simulator, the left atrial blood pool was segmented from a contrast enhanced magnetic resonance angiogram scan performed at St Thomas' Hospital \cite{sim_reproducibility_2019}. This segmentation was meshed using a marching cubes algorithm in CEMRGApp \cite{razeghi_cemrgapp_2020}, and the resulting surface was remeshed to a regular edge length of 0.3mm using mmgtools software \cite{dapogny_three-dimensional_2014}, corresponding to around 110,000 vertices, which is sufficient for simulation with the MMS model. This mesh can be found here \cite{roney_predicting_2022}, and is also included with our code \cite{sam_coveney_2022_7081857}.

\subsection*{Sensitivity Analysis}

To determine ERP(S1), the ERP value under an S1S2 protocol for S1 600~ms, and ERP(S2), the ERP value under an S1S2S3 protocol for S1 600~ms and S2 300~ms, we utilized a surrogate simulation: a strip of tissue with homogeneous parameters, paced from one end with the corresponding protocol, with activation measured in the strip centre \cite{coveney_bayesian_2021}. The strip simulation is set up to match the atrial simulation as closely as possible (space and time discretization, cell model time-step subdivision, numerical integration, etc). We obtain simulation results with an optimized Latin hyper-cube design of 500 parameter combinations in the parameter range explained above.

Variance-based sensitivity analysis was performed by fitting a General Additive Model (GAM) to model outputs, e.g. ERP(S1), as a function of a single model input, e.g. $APD_{max}$. The expectation of the GAM is then a line through a point-cloud of input-output pairs. The variance of this line (evaluated at a finite number of equally spaced locations) divided by the variance of the point-cloud gives an approximate sensitivity index of the input on that output \cite{strong_estimating_2016, wood_generalized_2017}. This method can be repeated for all inputs and all outputs. We implement GAMs using the LinearGAM function with 10 splines from the Python module PYGAM \cite{serven_dswahpygam_2018}. The sensitivity index of output y for input x can then be calculated as

\begin{lstlisting}
  gam = LinearGAM(n_splines=10).gridsearch(x, y)
  sen = gam.predict(x).var() / y.var()
\end{lstlisting}

\subsection*{Surrogate functions}

The map from EP parameters (inputs) to high dimensional tissue responses (outputs), such as restitution curves, has been modelled previously using Gaussian processes \cite{coveney_bayesian_2021}. Here, cubic polynomials in both $\theta_1 = \tau_{out}$ and $\theta_2 = APD_{max}$ were fit to corresponding values of $\text{ERP}_\text{S2}$ and $\text{ERP}_\text{S3}$, generated from an optimized Latin hyper-cube design of 100 values of $CV_{max}$, $\tau_{out}$ and $APD_{max}$, keeping $\tau_{in} = 0.05$~ms and $\tau_{open} = 120$~ms in order to produce ERP and APD values in a range observed in human atrial tissue \cite{bode_repolarization-excitability_2001}. $CV_{max}$ was varied for robustness, but has negligible effects on ERP, as confirmed by negligible fitting residuals. There is a discontinuity in $\text{ERP}_\text{S3}$ for parameter values producing $\text{ERP}_\text{S2} \approx 285$~ms, so data for $\text{ERP}_\text{S2}> 280$~ms were discarded before fitting these functions. We refer to these polynomial fits as `surrogate functions' for $\text{ERP}_\text{S2}$ and $\text{ERP}_\text{S3}$, denoted as $f_{1}(\theta_1, \theta_2)$ and $f_{2}(\theta_1, \theta_2)$ respectively, as they allow for determining ERP without running simulations.

\subsection*{Gaussian process priors}

We model the EP parameters fields, $\theta_l(\mathrm{x})$, as spatially correlated random fields defined on the atrial manifold, i.e. $\mathrm{x} \in \Omega$. We use Gaussian Process Manifold Interpolation (GPMI), a method  we proposed for defining Gaussian process distributions on manifolds \cite{coveney_gaussian_2020}. The approach  uses solutions $\left\{\lambda_k, \phi_k(\mathbf{x})\right\}$ of the Laplacian (Laplace-Beltrami) eigenproblem on the mesh \cite{solin_hilbert_2020}. 
Using GPMI allows us to represent fields on the atrium using a coordinate system that uses these eigenfunctions as a basis, enabling us to calibrate parameter fields on any given atrial manifold. The prior for each parameter field $\theta_l(\boldsymbol{x})$ can then be represented using the following probabilistic model, which uses the $K$ smallest eigenvalue solutions to the Laplacian eigenproblem:
\begin{gather}
    \theta_l(\boldsymbol{x}) = m_l + \alpha_l \sum_{k=1}^K (\boldsymbol{\eta_l})_k \sqrt{S\left(\sqrt{\lambda_k}, \rho_l\right) } \phi_k(\boldsymbol{x})   \label{eq:GP}  \\
    (\boldsymbol{\eta_l})_k \sim \mathcal{N}(0,1) \label{eq:eta} 
\end{gather}
\noindent
with hyperparameters mean $m_l$, amplitude $\alpha_l$, and lengthscale $\rho_l$, for $l=1,2$ and $k=1, \ldots, K$. The lengthscales determine the distance over which values are correlated, with larger lengthscale corresponding to smoother parameter fields. The units of the lengthscale hyperparameters are determined by the spatial units of the mesh on which the eigenproblem is solved. See below for details.
The hyperparameter vector $\boldsymbol{\eta_l}\in\mathbb{R}^K$ must be given a Gaussian prior in order for this model to approximate a Gaussian process. The function $S\left(\sqrt{\lambda_k}, \rho_l\right)$ is the spectral density corresponding to the choice of  covariance kernel, with the square root of the eigenvalue $\sqrt{\lambda_k}$ being the `frequency' argument to this function. In this work, we use the spectral density for the radial basis function (exponentiated quadratic) kernel, but other stationary kernels could be used.

It is possible and tractable to perform inference directly on the simulation mesh by solving the Laplacian eigenproblem on this mesh. But for convenience, a lower resolution mesh of 5000 vertices was used, with vertices that are a subset of the higher resolution mesh vertices. The lower resolution mesh is produced using a simulated annealing algorithm to optimally choose a subset of 5000 nodes before meshing these new nodes to form a new surface. The routines for this are found in the quLATi package \cite{coveney_samcoveneyqulati_2021}, and method details are given in our previous work \cite{coveney_gaussian_2020}. This lower resolution mesh allows for calculation of eigenfunctions with fewer computational resources, less data storage for eigenfunctions, and is convenient for plotting. Values can be easily transferred to the simulation mesh via interpolation (we use the `interpolate' function in the Python software Pyvista \cite{sullivan_pyvista_2019}). However, this `two-mesh' approach is entirely optional.
The units of lengthscale parameter $\rho$ in Eq \eqref{eq:GP} can be empirically related to geodesic distance by drawing many GP samples for a given kernel function on the mesh (using a lengthscale that allows for correlations to approach zero for some pairs of vertices since the mesh is finite), calculating the correlation between these samples at many pairs of vertices, and fitting (via least squares) the kernel function to correlation as a function of geodesic distance between the pairs of vertices. For the mesh in this work, 1 unit of kernel lengthscale corresponds to approximately $3.2$~mm.

\subsection*{Bayesian calibration}

Given ERP measurements at different locations $\mathbf{x}_i$ over the atrium, it is possible to calibrate the parameter fields $\theta_1(\mathbf{x}) \equiv \tau_{out}(\mathbf{x})$ and $\theta_2(\mathbf{x}) \equiv APD_{max}(\mathbf{x})$ by obtaining the posterior distribution of the hyperparameters in Eq. \eqref{eq:GP}. For convenience, we collect the hyperparameters into the vector $\boldsymbol{\psi} := (m_1, m_2, \alpha_1, \alpha_2, \rho_1, \rho_2, \boldsymbol{\eta_1}, \boldsymbol{\eta_2})$.
Defining the ERP measurements as $\mathbf{y}$, then we can write the Bayesian inference problem as:
\begin{gather}
    p(\boldsymbol{\psi} | \mathbf{y}) \propto p(\mathbf{y} | \boldsymbol{\psi}) p(\boldsymbol{\psi}) \label{eq:LH} \\
    p\left(\mathbf{y}|\boldsymbol{\psi}\right) := \prod_i p\left(y_1(\mathbf{x}_i) \;|\; \boldsymbol{\psi}\right) \prod_i p\left(y_2(\mathbf{x}_i) \;|\; \boldsymbol{\psi}\right)
\end{gather}
where $y_1(\mathbf{x}_i)$ and $y_2(\mathbf{x}_i)$ represent observations of $\text{ERP}_\text{S2}$ and $\text{ERP}_\text{S3}$ respectively. We assume that both types of ERP are measured at each location, but this is not a requirement as terms can just be replaced with $1$ if the corresponding measurement is not performed. 
Clinically, S1S2 protocols are performed by decreasing S2 by $\Delta$S2 until successful activation does not occur on the S2 beat. Therefore, observations of ERP are only observations of an interval in which ERP lies. The observation that each ERP value at a measurement location $\mathbf{x}_i$ lies between two S2 values $t_s$ and $t_{s+1}$ can be expressed in the following way (see Figure \ref{fig:workflow} for a graphical representation):
\begin{align}
    y_1(\mathbf{x}_i) := \text{ERP}_\text{S2} \in \left[t_3,\; t_4\right] \; \text{at} \; \mathbf{x}_i  \\
    y_2(\mathbf{x}_i) := \text{ERP}_\text{S3} \in \left[t_1,\; t_2\right] \; \text{at} \; \mathbf{x}_i
\end{align}

Observations can be linked to the hyperparameters $\boldsymbol{\psi}$ via the GP fields defined by Eq \eqref{eq:GP}, which determine the EP parameters at positions on the atrial mesh, and by the surrogate functions, which map these EP parameter values to ERP values:
\begin{gather}
    p\left(y_m(\mathbf{x}_i) \;|\; \boldsymbol{\psi}\right) = p\left(y_m(\mathbf{x}_i) \;|\; f_m(\mathbf{x}_i)\right) \\
    f_m(\mathbf{x}_i) := f_m\left(\theta_1(\mathbf{x}_i, \boldsymbol{\psi_1}),\; \theta_2(\mathbf{x}_i, \boldsymbol{\psi_2})\right)
\end{gather}
\noindent
where $\psi_1$ and $\psi_2$ represent partitions of $\boldsymbol{\psi}$ for each EP parameter field.

An S1S2 pacing protocol to determine ERP effectively measures the S2 interval in which ERP lies. Defining the lower bound of this interval by $I$ and the interval width by $\Delta S2$, the true likelihood is given by a truncated uniform, or `top-hat', distribution. In other words, $p(\text{ERP} \in [I, I + \Delta S2]  \;|\; \boldsymbol{\psi} )$ is equal $1$ if $\boldsymbol{\psi}$ produces ERP in the specified interval, and $0$ otherwise.
The surrogate functions $f_m(\theta_1, \theta_2)$ can be used to predict ERP from the EP parameters, which are determined by the GP fields $\theta_l(\mathbf{x})$ depending on the hyperparameters $\psi$.

However, it is more convenient to work with an approximation to this top-hat distribution, which we previously derived for use with ERP measurements \cite{coveney_bayesian_2021}. This top-hat likelihood can be approximated by dividing the interval into $N$ sub-intervals, with a normal distribution $\mathcal{N}(c_i, s)$ centered on each sub-interval $c_i = I + (i - 1/2)\Delta S2 / N$ with standard deviation equal to the sub-interval width $s = \Delta S2 / N$. We choose $N = \Delta S2$, such that $s = 1$. For an observation $y_1(\mathbf{x^*}) := \text{ERP}_\text{S2} \in \left[I,\; I + \Delta S2\right]$ (and similarly for $\text{ERP}_\text{S3}$) the likelihood can be approximated as:

\begin{align}
    p(y_1 \in [I, I + \Delta S2]  \;|\; \boldsymbol{\psi} ) = \frac{1}{N} \sum_{i=1}^{N} \frac{1}{\sqrt{2 \pi s^2}} \exp{\left(- \frac{ ( f_{1}(\boldsymbol{\theta}(\mathbf{x^*})) - c_i )^2 }{2 s^2} \right)}
\end{align}

The shape of this likelihood function is a top-hat with smoothed sides and no discontinuities, such that the likelihood is approximately constant in the interval but rapidly falls to zero near the interval edges. This approximate top-hat distribution has infinite support, allowing gradient-based MCMC to be performed. For the log-likelihood, the readily available \texttt{logsumexp} function is used to prevent numerical underflow (this function is available in STAN \cite{team_stan_2021, riddell_pystan_2021}, which is used for MCMC for this work). Note that this approximate top-hat distribution integrates to 1, rather than having approximately constant value 1 (in the interval), but constant factors do not matter for MCMC so we retain this form for simplicity. Also note that with a small adjustment this likelihood can be used with a Gaussian process surrogate function that predicts mean and variance \cite{coveney_bayesian_2021}.

%
%
%
%
%
%
%
We use STAN (via PyStan) \cite{team_stan_2021, riddell_pystan_2021} to perform Hamiltonian MCMC, which yields samples from the posterior distribution of the parameters $\boldsymbol{\psi}$. See Results for details. We can use these samples to calculate samples of the EP fields over the entire atrium using Eq. \eqref{eq:GP}, from which ERP samples can be calculated using the surrogate functions. We modify equation \eqref{eq:GP} by replacing $\alpha_l$ with $\alpha_l / |\Phi_1|$ (where $\Phi_1(\mathbf{x})$ is constant), which assists with defining priors for $\alpha_l$. We used the following priors on the hyperparameters: $\rho_l  \sim  \text{InvGamma}(1.01, 20)$, $m_l  \sim  \text{Uniform}(-\infty, +\infty)$, and $\alpha_{l}  \sim  \text{InvGamma}(1, 5)$. We found that these priors consistently allow for recoverability of both EP parameter fields. Eq \eqref{eq:eta} gives the prior for the remaining hyperparameters.

\subsection*{Parameter samples}

To generate `ground truth' parameter fields, we draw samples from a Gaussian process defined by Eq \eqref{eq:GP} with Matern 5/2 spectral density function using 256 eigenfunctions. We set parameters $m = 0$ and $\alpha = 1$, and $\rho$ is set to values explained in Results and below. The generated samples for $\tau_{out}$ and $APD_{max}$ are then scaled and offset into the full allowable parameters ranges. The same operation is performed for $CV_{max}$, which is needed for atrial simulations.

Certain combinations of $\tau_{out}$ and $APD_{max}$ correspond to regions of parameter space that produce unrealistic ERP. This is handled for both `ground truth' samples and posterior samples by identifying mesh nodes where the parameters produce values $ERP(S1) > 280~ms$. The parameter values at these nodes are then replaced by a weighted average of parameter values at other nodes with acceptable ERP values, weighted by $1/d_{BH}^4$ where $d_{BH}$ is biharmonic distance. Biharmonic distance, calculated from the Laplacian eigenvalues and eigenfunctions, is significantly cheaper to calculate than geodesic distance, and avoids topological issues from using Euclidean distance \cite{lipman_biharmonic_2010} to interpolate values on a manifold. This procedure allows to constrain parameter samples efficiently and effectively, and is far simpler than attempting to encode such constraints into MCMC.

\subsection*{Synthetic experiment}
%
We created ground truth parameter fields for $\tau_{out}$ and $APD_{max}$ in order to verify our calibration approach (see above for details). ERP values were calculated using the surrogate functions. For ERP measurements, we generate a design of measurement locations using an optimized `maximin' hypercube design, excluding mesh sites within 0.6~cm of mesh boundaries as potential sites for these measurements since clinical measurements are unlikely to sample these regions. The resolution of the S1S2 and S1S2S3 protocol set to values specified in Results. A lengthscale of 20 was used for the example shown in Figures \ref{fig:parameters} -- \ref{fig:APD}.

APD values were obtained using the atrial simulator (simulations of the mono-domain equation with the mMS model using the software \emph{openCARP}). Tissue was paced for 8 beats from near the coronary sinus, and depolarization and repolarization were measured on the final beat.  A spatially varying $CV_{max}$ field was generated for use in this simulation, and $\tau_{in}$ and $\tau_{open}$ were fixed as described above.
Simulations were run either for ground truth parameter fields or for predicted parameter fields resulting from calibration.
Note that the parameters described in this manuscript were transformed back into the original parameters for the mMS model for running simulations in \emph{openCARP}. The diffusion time-step was 0.1~ms and the ionic current time-step was 0.02~ms. The mMS action potential is normalized to have minimum 0 and maximum 1. Activation (depolarization) was measured when $V_m$ reached 0.7 on upstroke, and recovery (repolarization) was measured when $V_m$ fell to 0.8 ($\text{APD}_{20}$), 0.7, ($\text{APD}_{30}$), 0.5 ($\text{APD}_{50}$), and 0.1 ($\text{APD}_{90}$). APD values are the time between activation and recovery. Simulations results for $\text{APD}_{20}$ and $\text{APD}_{90}$ are given in Results.

\section*{Acknowledgements}

This work was funded by grants from the UK Engineering and Physical Sciences Research Council (EP/P010741/1, EP/P01268X/1). CHR acknowledges a Medical Research Council Skills Development Fellowship (MR/S015086/1).

\section*{Author contributions statement}

SC conceived and implemented the work. CHR and CC contributed to atrial simulation scripts. RDW and JEO advised on sensitivity analysis and calibration. SDN and RHC provided guidance on cardiac modelling and the clinical setting. All authors contributed to the manuscript.

\section*{Data availability}

Our code and example mesh are available in a Zenodo repository \cite{sam_coveney_2022_7081857}.

\section*{Additional information}

The authors declare no competing interests.

\bibliography{bibliography}

\end{document}